\RequirePackage{ifpdf}
\documentclass[11pt,a4paper]{article}
\pdfoutput=1
\usepackage{jcappub}
\usepackage{amsmath,amssymb,epsf}
\usepackage{graphicx,wrapfig,color}
\usepackage{subfig}
\usepackage{setspace}
\usepackage{makecell}
\usepackage{tabularx}
\usepackage{booktabs}
\usepackage[tablename=Table]{caption}
\usepackage{float}
\usepackage{color}
\definecolor{cblue}{RGB}{100,5,255}
\definecolor{cred}{RGB}{255,50,40} 
\definecolor{cgreen}{RGB}{255,0,255} 
\definecolor{corange}{RGB}{250,200,40} 
     
\def\bel#1{\begin{equation} \label{#1}}

\def\be{\begin{equation}}
\def\ee{\end{equation}}
\def\bea{\begin{eqnarray}}
\def\eea{\end{eqnarray}}

\def\ltap{\ \raise.3ex\hbox{$<$\kern-.75em\lower1ex\hbox{$\sim$}}\ }
\def\gtap{\ \raise.3ex\hbox{$>$\kern-.75em\lower1ex\hbox{$\sim$}}\ }
\def\gl{\ \raise.5ex\hbox{$>$}\kern-.8em\lower.5ex\hbox{$<$}\ }
\def\roughly#1{\raise.3ex\hbox{$#1$\kern-.75em\lower1ex\hbox{$\sim$}}}

\newcommand{\comments}[1]{}

\newcounter{oldcounter}
\addtocounter{equation}{1}
\title{Constraining Warm Inflation with CMB data}
\author[a]{Mar Bastero-Gil,}
\author[b, c]{Sukannya Bhattacharya,}
\author[b,c]{Koushik Dutta}
\author[b]{Mayukh Raj Gangopadhyay}
\affiliation[a]{\small Departamento de F\'{\i}sica Te\'orica y del Cosmos, Universidad de Granada,\\ Granada, 18071 Spain}
\affiliation[b]{\small Theory Division, Saha Institute of Nuclear Physics, 1/AF Bidhan Nagar,\\ Kolkata, 700064 India }
\affiliation[c]{\small Homi Bhabha National Institute, Training School Complex, Anushaktinagar,\\ Mumbai,
400094 India}
\emailAdd{mbg@ugr.es} \emailAdd{sukannya.bhattacharya@saha.ac.in} \emailAdd{koushik.dutta@saha.ac.in} \emailAdd{mayukh.raj@saha.ac.in}

\abstract{
We confront the warm inflation observational predictions directly with the latest CMB data. We focus on a linear temperature $(T)$ dissipative coefficient combined with the simplest model of inflation, a quartic chaotic potential. Although excluded in its standard cold inflation version, dissipation reduces the tensor-to-scalar ratio and brings the quartic chaotic model within the observable allowed range. We will use the CosmoMC package to derive constraints on the model parameters: the combination of coupling constants giving rise to dissipation, the effective number of relativistic degrees of freedom contributing to the thermal bath, and the quartic coupling in the inflaton potential. We do not assume a priori a power-law primordial spectrum, neither we fix the no. of e-folds at the horizon exit. The relation between the no. of e-folds and the comoving scale at horizon crossing is derived from the dynamics, depending on the parameters of the model, which allows us to obtain the $k$-dependent primordial power spectrum. We study the two possibilities considered in the literature for the spectrum, with the inflaton fluctuations having a thermal or a non-thermal origin, and discuss the ability of the data to constraint the model parameters. 
}

\begin{document}

\maketitle

\section{Introduction}

Cosmological observations are in very good agreement with an universe that is expanding, spatially flat, homogeneous and isotropic on large scales, and where the large scale structure originated from primordial perturbations with a nearly gaussian and scale-invariant spectrum \cite{wmap, planck2013cosmo, planck2015cosmo}. On the theoretical side, the paradigm of slow-roll inflation, a period of accelerated expansion in the early history of the universe, predicts such a primordial spectrum starting with the fluctuations of the inflaton field \cite{inflation, starobinsky, inflation2}. In the standard paradigm, inflaton quantum fluctuations are stretched out of the horizon due to the expansion, and transferred to the curvature perturbation with constant amplitude spectrum on super-horizon scales. 

However because the Cosmic Microwave Background (CMB) radiation spectrum can be well explained with just a power-law primordial spectrum \cite{planck2015}, we have information on the amplitude and the spectral index, but so far not much more than that. For example there is as yet no detection of a primordial tensor component, which translates in an upper limit on the energy scale at which inflation took place. For single field models, where the dynamics during inflation is just controlled by a potential energy density, this seems to favour plateau-like potentials \cite{starobinsky, plateau} or small field models \cite{small1, small2},  as opposed to large field models \cite{largeinf, McAllister:2008hb} for which the potential energy (and the tensor-to-scalar ratio) is larger.    

This is the situation in the standard scenario of slow-roll inflation, which we can call ``Cold Inflation'' (CI), given that any other component of the energy density, and in particular radiation, will be quickly redshifted away even if present initially. However, inflation has to be followed by a radiation dominated period to allow for the synthesis of primordial nuclei (BBN), which requires the conversion of the inflaton energy density into radiation during the so-called (p)reheating period \cite{reheating, preheating}. This neccessarily implies interactions among the inflaton field and other light degrees of freedom, which may already play a role during inflation. Thus, the transfer of energy between the inflaton and radiation may start during inflation. This is the warm inflation (WI) scenario \cite{Berera:1995wh}, where the energy transfer translates into an extra friction or dissipative term $\Upsilon$ in the field EOM. The extra friction therefore favours slow-roll inflation, slowing down even further the evolution of the inflaton. Inflation can last for longer, and the relevant part of inflation when the primordial spectrum originates can happen at a smaller energy density value, which gives rise to a suppressed tensor-to-scalar ratio. The nature of the primordial spectrum can be completely different in WI due to the influence of the thermal bath fluctuations on the inflatons, such that the fluctuations will have now a thermal origin \cite{RS13, BBMR14}. 

The specific functional form of the dissipative coefficient $\Upsilon$ with the inflaton field $\phi$ and the temperature $T$ of the plasma will depend on the pattern of the inflaton interactions with other degrees of freedom \cite{Moss:2006gt, BasteroGil:2012cm}. Dissipation for example may originate from the coupling of the inflaton and a heavy field mediator,  which in turn decays into relativistic particles. This pattern does not introduce any thermal correction in the inflaton potential, the contribution of the mediators with mass $m_\chi >T$ being Boltzmann suppressed. Therefore, it does easily overcome the main difficulty faced originally to build viable warm inflation models \cite{highT}, i.e, preserving the required flatness of the potential to allow slow-roll inflation. However the dissipative coefficient is only power law suppressed and one gets $\Upsilon \propto T (T/m_\chi)^\alpha$, where the power $\alpha$ of the ratio $T/m_\chi \lesssim 1 $ depends on the bosonic/fermionic nature of the mediator and its decay products. Although it can give rise to viable models of inflation consistent with observations\footnote{See Refs. \cite{others1,others4, others5} for consistent models of warm inflation with other dissipative coefficients.} \cite{importance, nico}, it typically requires a large number of mediator fields for the effects of dissipation to be sizeable.   

When the mediators are light, for example fermions directly coupled to the inflaton, one has to check that the induced thermal corrections to the inflaton potential are under control, while still having strong enough interactions to allow the thermalization of the light degrees of freedom, and giving rise to enough dissipation. A scenario fulfilling these conditions has been recently proposed in \cite{BBRR16}. In the same spirit as ``Little Higgs'' models, the inflaton is a pseudo-Nambu Goldstone boson (PNGB) of a broken gauge symmetry, its $T=0$ potential being protected against large radiative corrections by the symmetry. Similarly, in order to avoid large $T$ corrections due to the light fermions, a discrete (exchange) symmetry is imposed in the inflaton and fermionic sectors. This ensures that the leading field dependent thermal mass correction cancels out, leaving only the subleading $T$-dependent logarithmic one. This leads to a dissipative coefficient just linear in $T$, and to enough dissipation without the need of large no. of fields.  

Given the possibilities for inflationary model building open up by the combination of symmetries and interactions/dissipation with a linear $T$ coefficient, it is worth exploring the observational consequences and in particular confront directly the model with CMB data. We will use the CosmoMC package to perform a multi-dimensional Markov Chain Monte Carlo (MCMC) analyses and derive constraints directly on the model parameters. We will focus on the simpler potential, the quartic chaotic potential $\lambda \phi^4$ model, which although excluded in its CI version\footnote{Although it has been pointed out that 1-loop radiative corrections due to the interactions with other species can render the potential flat enough to lower the tensor-to-scalar ratio below the observable upper limit \cite{NeferSenoguz:2008nn, Enqvist:2013eua, Ballesteros:2015noa}.}, it is compatible with data once the effects of the interactions are included \cite{BBRR16}. Therefore the inflationary dynamics and the spectrum will be given by three parameters: the coupling $\lambda$ in the inflaton potential; the combination of couplings $C_T$ leading to linear dissipation $\Upsilon=C_T T$; and the effective no. of light degrees of freedom contributing to the thermal bath, $g_*$. The amplitude of the primordial spectrum and its scale dependence can be derived as a function of these parameters, and the prediction compared directly with the data without the need a priori of assuming a power-law parametrization. The scale dependence, given in terms of the comoving $k$ scale at which perturbations exit the horizon during inflation, can be related to the no. of efolds $N$ but implies some assumption about the (p)reheating period \cite{Easther10}. Typically one gets $N \in [50,60]$ for the no. of efolds at which the largest observable scale crosses the horizon, and predictions are quoted varying $N$ in this range. Our choice of the potential allows us to avoid this uncertainty in the predictions, given that the quartic potential will behave as radiation once inflation ends.      

Recently in Ref.~\cite{Benetti} the authors performed a thorough analyses of the different popular models of inflation 
in both the low and high temperature regimes. The low temperature regime is defined by the cubic dissipative coefficient whereas the high temperature regime is described by a linear dissipative coefficient. In this work our focus is on this high temperature regime which has been first described in \cite{BBRR16}. Ref.~\cite{Benetti} did a statistical analyses of the models using CMB data, and showed the viability of the scenario for different models which are excluded in CI scenario from present and (projected) future observations. Their work, besides the predictions for the spectral index and the tensor-to-scalar ratio, is largely motivated to accommodate the latest observed values of the running and the running of the running of the spectral index\footnote{General consistency relations for WI including the scale dependence of the spectral index were first considered in \cite{Bartrum:2013oka}.}. Instead, in this paper, we have chosen to work only with a particular, well motivated model as the quartic chaotic model, and study how well the data can constrain the parameters of the model. Some of the main differences in approach between Ref.~\cite{Benetti} and our work are that, firstly, they fixed the potential parameters with the observable value of the amplitude of the spectrum, and $g_*$ depending on the model; while we have kept both the model parameter ($\lambda$) and $g_*$ as variables. Secondly, they chose to work with a fixed no. of $e$-folds, $N=55$, while we vary the number of $e$-folds of infation because of its implicit dependence on the model dynamics. Moreover, they focus on the  dependence with respect to the dissipative ratio, $Q=\Upsilon/(3H)$, i.e, the dissipative coefficient normalized to the Hubble parameter $H$. This would be equivalent to our choice of parameter $C_T$. And more crucially, they always consider the inflaton to be included in the thermal bath which is equivalent to the thermal case discussed in our work, but we additionally analysed the case where the inflaton does not have a thermal distribution (non-thermal case). 

The plan of the paper is as follows. In Sect.~\ref{basic} we describe the basic of the warm inflation dynamics, and the validity of slow-roll approximation during inflation. In Sect.~\ref{primordial} we give the expression for the primordial spectrum as a function of the parameters of the model, and how to get its scale dependence. Using the analytical expressions, we explore the parameter dependence of the predictions in Sect.~\ref{back_ana}. In Sect.~\ref{methodology} we describe the technical details of the analyses done with the MCMC and the CMB data, while in Sect.~\ref{results} we present the main results. Finally a summary and the conclusions of this work are given in Sect.~\ref{conclusions}.

\section{Basics of Warm Inflation}
\label{basic}
In warm inflation, the transfer of energy  between the inflaton scalar field $\phi$ and the plasma leads to an additional friction term in the inflaton equation of motion \cite{Berera:1995wh,Berera:2008ar},  described by the damping coefficient $\Upsilon(\phi,T)$. 
When dissipation leads to the production of  light degrees of freedom which 
thermalize in less than a Hubble time, then a radiation fluid
$\rho_r$ is produced, continually replenished by
the effective decay of the inflaton field.  
The background evolution equations for the inflaton-radiation system are 
given by: 
\begin{eqnarray}
\ddot\phi+(3H+\Upsilon)\dot\phi+V_{,\phi}&=&0, \label{ddotphi}\\ 
\dot\rho_r+4H \rho_r &=& \Upsilon\,\dot\phi^2 \,,
\end{eqnarray}
where a ``dot'' denotes time derivatives, $V_{,\phi}=d V/d \phi$, $V$ is the potential energy density, and $H$ the Hubble parameter: 
\be
3H^2= \frac{\rho}{M_P^2} \,,\label{hubble}
\ee
$\rho$ being the total energy density of both field and radiation, and $M_P$ is the reduced Planck mass. 
The radiation fluid is made of $g_*$ relativistic degrees of freedom at temperature $T$, with:
\be
\rho_r = \frac{\pi^2}{30}g_* T^4 = C_R T^4 \,.\label{deftemp}
\ee
Prolonged inflation requires the slow-roll conditions $|\epsilon_X|\ll
1$, where $\epsilon_X = - d \ln X/Hdt$, and $X$ is any of the
background field quantities. The
background equations at leading order in the slow-roll approximation of small $\epsilon_X$ become
\begin{eqnarray}
3H(1+Q)\dot\phi&\simeq&-V_{,\phi}\,, \label{sl1}\\ 
4 \rho_r&\simeq&3Q\dot\phi^2\,, \label{sl2}\\ 
3H^2&\simeq& \frac{V}{M_P^2}\,, \label{sl3}
\end{eqnarray}
where $Q=\Upsilon/(3H)$ is the dissipative ratio.

We will consider a linear $T$ dissipative coefficient like in Ref. \cite{BBRR16}. Dissipation comes from the coupling of the inflaton field to a pair of fermions with coupling $g$, while the latter interacting with a light scalar field with coupling $h$. We stress that the calculation of the dissipative coefficient is done in the adiabatic and quasi-equilibrium approximation, which impose some restrictions on the values of the parameters. First, once the inflaton excites the fermions to which it directly couples, they decay into scalars which have to thermalize in less than a Hubble time, i.e., the decay rate must be larger than $H$. In addition we require $T > H$, such that dissipation can be computed in the limit of flat spacetime with the standard tools of Thermal Quantum Field Theory \cite{TQFT}. Under those restrictions, the dissipative coefficient is then given by $\Upsilon = C_T T $, with $C_T$ being a function of the couplings: 
\be 
C_T\simeq \frac{3 g^2/h^2}{1-0.34 \ln h} \label{CT}\,.
\ee

For the inflaton potential, we work with the single-field chaotic quartic potential, 
\be
V(\phi)= \lambda \phi^4\,,
\ee
which is not excluded by observations once dissipation is taken into account \cite{importance, BBRR16}: the extra friction slows down the motion and effectively  ``flattens'' the potential seen by the inflaton, and therefore it tends to lower the predicted tensor-to-scalar ratio.  The slow-roll parameters are given by:
\bea
\epsilon_\phi&=&\frac{M_p^2}{2}\left(\frac{V_\phi}{V}\right)^2= 8\left(\frac{M_p}{\phi}\right)^2\,,\\
\eta_\phi&=&M_p^2\left(\frac{V_{\phi\phi}}{V}\right)^2= 12
  \left(\frac{M_p}{\phi}\right)^2\,,\\
\sigma_\phi&=&M_p^2\left(\frac{V_\phi/\phi}{V}\right)= 4
\left(\frac{M_p}{\phi}\right)^2\,.
\eea
Notice that, given the extra friction term $\Upsilon$, to have slow-roll
inflation we now require:
\be
\epsilon_\phi < 1+Q\,,~~~~\eta_\phi <(1+Q)\,,~~~\sigma_\phi < (1+Q)\,.
\ee
From the slow-roll equations (\ref{sl1}-\ref{sl3}) one can get the relation between $Q$ and $\phi$:
\bea
Q^3(1+Q)^2 &=& \frac{4}{9}\left(\frac{ C_T^4}{ C_R \lambda} \right)\left(\frac{m_P}{\phi}\right)^6~.
\label{Qphi}
\eea
Similarly, one can write directly the evolution equation for the
dissipative ratio $Q$, with respect to the no. of e-folds $dN = H dt$:
\be
\frac{dQ}{d N} \simeq \frac{Q}{3 + 5Q} \left( 6 \epsilon_\phi - 2
\eta_\phi \right) \,,
\ee
which for the quartic potential, and using Eq. (\ref{Qphi}) reduces to:
\be
\frac{dQ}{d N} \simeq C_Q \frac{ Q^2 (1+Q)^{2/3}}{3 + 5Q} \,,
\label{dQNe}
\ee
where:
\be
C_Q= 24 \left(\frac{4 C_T^4}{9 C_R \lambda}\right)^{-1/3} \,.
\ee
Eq. (\ref{dQNe}) can be integrated in terms of hypergeommetric functions:
\bea
C_Q N &=& f(Q_e) - f(Q_*) \,, \nonumber \\
f(x)&=& -3 \left( \frac{(1+x)^{1/3}}{x} + \frac{3}{x^{5/3}}~_2 F_1
[2/3, 2/3, 5/3, -1/x] \right) \,. \label{QNe}
\eea
By $Q_*$ we denote the value at horizon crossing of observable modes at CMB at $N$ e-folds before the end of inflation, and by $Q_e$ the value of the dissipative ratio at the end of inflation. We take the condition $\eta_\phi = 1+ Q_e$ signaling the end of inflation, and using this condition in Eq. (\ref{Qphi}) we have:
\be
\frac{Q_e^3}{(1 + Q_e)} = \frac{C_T^4}{2^4\times 3^5 \times C_R \lambda}\,. \label{Qend}
\ee
Given a value of $N$, Eq. (\ref{QNe}) can be inverted (numerically) to get the value of $Q_*$ (and then $\phi_*$),  needed to evaluate the amplitude of the primordial spectrum. We will revise this in the next section.

\section{Warm inflation: primordial spectrum}
\label{primordial}

The general expression for the amplitude of the primordial spectrum, independent of the nature of the dissipative coefficient, is given by \cite{RS13,BBMR14}:
\be 
P_{\mathcal{R}} = ( P_{\mathcal{R},\,diss} + P_{\mathcal{R},\,vac} )
=  \left( \frac{ H_* }{\dot
  \phi_*}\right)^2 \left(\frac{H_*}{2 \pi}\right)^2
\left[
  \frac{T_*}{H_*} \frac{ 2\pi Q_*}{\sqrt{ 1 + 4 \pi Q_*/3 }} + 1+2
       {\cal N}_* \right]\,,  \label{PR}
\ee 
where all variables are evaluated at horizon crossing. 
The first term is the contribution due to the effect of 
dissipation on the inflaton fluctuations. In the limit of no dissipation, we would recover the standard expression for the primordial spectrum, but allowing the inflaton fluctuations to be in a statistical state other than the vacuum; 
for example being in a thermal excited state with ${\cal N}_* = n_{BE}(a_*H_*)=(e^{H_*/T*}-1)^{-1}$. The standard Bunch-Davies vacuum is given by ${\cal N}_*=0$. The latter case will be called in the following as ``non-thermal'' inflaton fluctuations, while we use ``thermal'' for the ${\cal N}_* = n_{BE}(a_*H_*)$ case. 
In \cite{BBMR14} it has been checked that indeed the analytic
solution of Eq. (\ref{PR}) reproduces the spectrum of warm inflation upto
values $Q_* \lesssim 0.1$, by numerically integrating the equations for the fluctuations.  

For larger dissipation at horizon crossing,
the spectrum gets enhanced due to the coupling between inflation and radiation fluctuations. This effect depends on $Q_*$, and can be accounted for by 
multiplying the spectrum in Eq. (\ref{PR}) by a function $G[Q_*]$ \cite{BBMR14},
\be
G[Q_*] \simeq 1 + 0.0185Q_*^{2.315}+ 0.335 Q_*^{1.364} \label{GQ}\,.
\ee
This parametrization however depends on both the inflaton potential and 
the $T$ dependence of the dissipative parameter. We quote in Eq. (\ref{GQ})
the values obtained for a quartic potential with a linear $T$ dependent $Q$
\cite{BBRR16}. 

In the above expression for the spectrum, one
can replace $T_*/H_*$ by $3 Q_*/C_T$ and the field dependence (in $H_*$ and
$\dot \phi_*$) in terms of $Q_*$ using Eq. (\ref{Qphi}):
\be
P_{\cal R} = \frac{C_T^4}{4 \pi^2\times 36  C_R} Q_*^{-3}
\left[
  \frac{3 Q_*}{C_T} \frac{ 2\pi Q_*}{\sqrt{ 1 + 4 \pi Q_*/3 }} + 1+2
       {\cal N}_* \right]\times G[Q_*]
\,. \label{PRQ*}
\ee
Therefore, the spectrum is given implicitly as a function of the
parameters of the model, $C_T$, $C_R(g_*)$ and $\lambda$, and the no. of
e-folds $N$ through Eq. (\ref{QNe}). In the case of having
a  thermal spectrum for the inflaton fluctuations, we also have:
\be
1 + 2 {\cal N}_* = \coth \frac{H_*}{2 T_*} = \coth \frac{C_T}{6 Q_*} \,.
\label{coth}
\ee

The no. of e-folds can be related to the scale at which
the fluctuation exits the horizon $k=a_* H_*$, so that finally we have
the spectrum as a function of the comoving scale $k$. 
However, the relation between $N$ and $k$ depends on the details of
reheating  \cite{liddleleach, Martin06, Martin10, Easther10}, the period between the end of inflation and a radiation dominated universe \cite{Dutta:2014tya}. 
Modeling our ignorance about reheating with an effective equation of state $\tilde w$, the relation between the no. of efolds and the comoving wavenumber is given by \cite{Easther10}:
\bea
N(k) &=& 56.12 - \ln \frac{k}{k_0} + \frac{1}{3(1+\tilde w)} \ln
\frac{2}{3} +\ln \frac{V_k^{1/2}}{V_{end}^{1/2}} + \frac{1-3 \tilde w}{3(1+ \tilde w)}\ln \frac{\rho_{RH}^{1/4}}{V_{end}^{1/4}} 
+ \ln \frac{V_{end}^{1/4}}{10^{16}\,{\rm GeV}} \,, \label{Nkw}
\eea
where $k_0=0.05\, {\rm Mpc}^{-1}$ is the pivot scale for Planck, $V_k$ and $V_{end}$ the potential values at the end and $N(k)$ e-folds before the end of inflation respectively, and $\rho_{RH}$ the energy density at the end of reheating when the universe becomes radiation dominated.
Typically the no. of efolds at which the largest observable scale leaves the horizon lies between $50-60$. But this intrinsic uncertainty in the inflationary predictions on the no. of e-folds is avoided in warm inflation with a quartic potential. In this case the dissipative ratio $Q$ increases during inflation, such that the radiation by the end becomes comparable to the inflaton energy density (signalling also the end of inflation). And for a quartic potential, once the field starts oscillating around the minimum of the potential, the average energy density behaves as radiation. It does not matter when the inflaton finally decays after inflation, because the universe is already radiation dominated. This is equivalent to having instant reheating (i.e., and instant
transition between inflation and the radiation dominated epoch), with $\tilde w=1/3$ and $\rho_{RH}=V_{end}$ in Eq. (\ref{Nkw}): 
\bea
N(k) &=& 56.02 - \ln \frac{k}{k_0}
+\ln \frac{V_k^{1/2}}{V_{end}^{1/2}}
+ \ln \frac{V_{end}^{1/4}}{10^{16}\,{\rm GeV}} \,. 
\label{Nk}
\eea
Therefore, instead of taking a certain $N$ interval to derive the observable predictions of the model, like in other studies \cite{Benetti}, we will compute directly the $k$-dependent power spectrum using Eq. (\ref{Nk}). The value of the potential at the end of inflation can be obtained with the value of $Q_e$ in Eq. (\ref{Qend}) and
\be
12 \frac{m_P^2}{\phi_e^2} = \frac{12 \lambda }{V_{end}^{1/2}}=1 + Q_e\,.
\ee
And for the ratio $(V_k/V_{end})^{1/2}=(\phi_k/\phi_e)^2$ we have:
\be
\left(\frac{\phi_k}{\phi_e} \right)^2= \frac{Q_e (1+ Q_e)^{2/3}}{ Q_k (1+
  Q_k)^{2/3}} \,.
\ee
Through the field dependence in $V_k$ and $V_{end}$, the relation
between the scale $k$ and the no. of efolds depends on the parameters
of the model $C_T$, $\lambda$ and $C_R(g_*)$. 

The primordial tensor spectrum is not affected by dissipation, so we have the standard prediction:
\be
P_T = 8 \left( \frac{H_*}{2 \pi m_p}\right)^2 \,,
\ee
which for a quartic potential is just given by:
\be
P_T = \frac{8 \lambda}{4 \pi^2} \left( \frac{\phi_*}{m_p}\right)^4 
=
\frac{8 \lambda^{1/3}}{4 \pi^2} \left( \frac{4 C_T^4}{9 C_R}\right)^{2/3} \frac{1}{Q_*^2 (1+Q_*)^{2/3}} \,,
\label{PTparam}
\ee
where we have used Eq. (\ref{Qphi}). 
Finally, the tensor-to-scalar-ratio in terms of  the parameters of the model (and $Q_*$) is given by:
\be
r=\frac{P_T}{P_{\cal R}}= 32 
\left(\frac{16 C_T^4}{9\lambda C_R}\right)^{-1/3} Q_*^3
\left[
  \frac{3 Q_*}{C_T} \frac{ 2\pi Q_*}{\sqrt{ 1 + 4 \pi Q_*/3 }} + 1+2
       {\cal N}_* \right]^{-1}\times G[Q_*]^{-1}\,. 
\label{tensorrat}
\ee
In the next section, we will calculate the scalar and tensor power spectrum as a function of $k$, and will determine its parameter dependence followed by the best fit parameter estimation in the following section. 

\section{Analysing parameter dependence on observables}
\paragraph*{} 
\label{back_ana}
In this section we will analyse parameter dependence of warm inflation model on inflationary observables. In particular, we will consider the scalar spectral index $n_s $ and the scalar amplitude $A_s$ as observables, which are functions of the parameters $C_T$, $\lambda$ and $g_*$. For the choices of the parameters, the upper bound on the tensor amplitude $r$ would be trivially satisfied, and for that purpose we do take it as a constraint in this section. But, for parameter estimation in the next section, the tensor amplitude would be incorporated accordingly. For our consideration, we will see that the running of the spectral index would be very small (being consistent with recent observations \cite{Benetti}, \cite{Silkalphas}), and we will not consider it as a constraining observable. The analysis of this section would be useful in determining the range of parameters as priors for the CosmoMC simulations and parameter estimation later. 
 
Our first goal is to calculate the scalar amplitude given by Eq.~\eqref{PRQ*} as a function of comoving wavenumber $k$. Other than the model parameters, the expression depends on the dissipative ratio $Q$ that needs to be calculated at horizon crossing for each wavenumber. The $k$ dependence of the scalar amplitude is implicit via its dependence in $Q$. For a particular set of $C_T$, $\lambda$ and $g_*$, the value $Q_e$ (the value at the end of inflation) is determined from Eq. (\ref{Qend}), and using Eq.~(\ref{Qphi}) $\phi_e$ can be calculated. Note that the Eq.~\eqref{Qphi} gives one to one relation between $\phi$ and $Q$. On the other hand, using Eq.~\eqref{QNe} and Eq.~\eqref{Nk}, $Q$ can be solved as a function of $k$. But instead of inverting the hypergeometric function of Eq.~\eqref{QNe}, we solve Eq.~\eqref{dQNe} and Eq.~\eqref{Nk} iterataively (numerically) and find $Q(k)$. We plug it in Eq.~\eqref{PRQ*} to find the scalar amplitude as a function of $k/k_0$ for both the non-thermal ($\mathcal{N}_*=0$) and thermal ($\mathcal{N}_*\neq0$) cases. We follow the same procedure in calculating the tensor amplitude from Eq.~\eqref{PTparam}. This algorithm is incorporated in the CAMB code \cite{cambb} in the form of a subroutine in calculating the $C_l$ s for the two-point correlation functions. The pivot scale is taken at usual $k = 0.05$ Mpc$^{-1}$ throughout the analysis. 
 



  


\begin{figure}[H]
  \centering
  \subfloat[]{\includegraphics[width=0.50\textwidth]{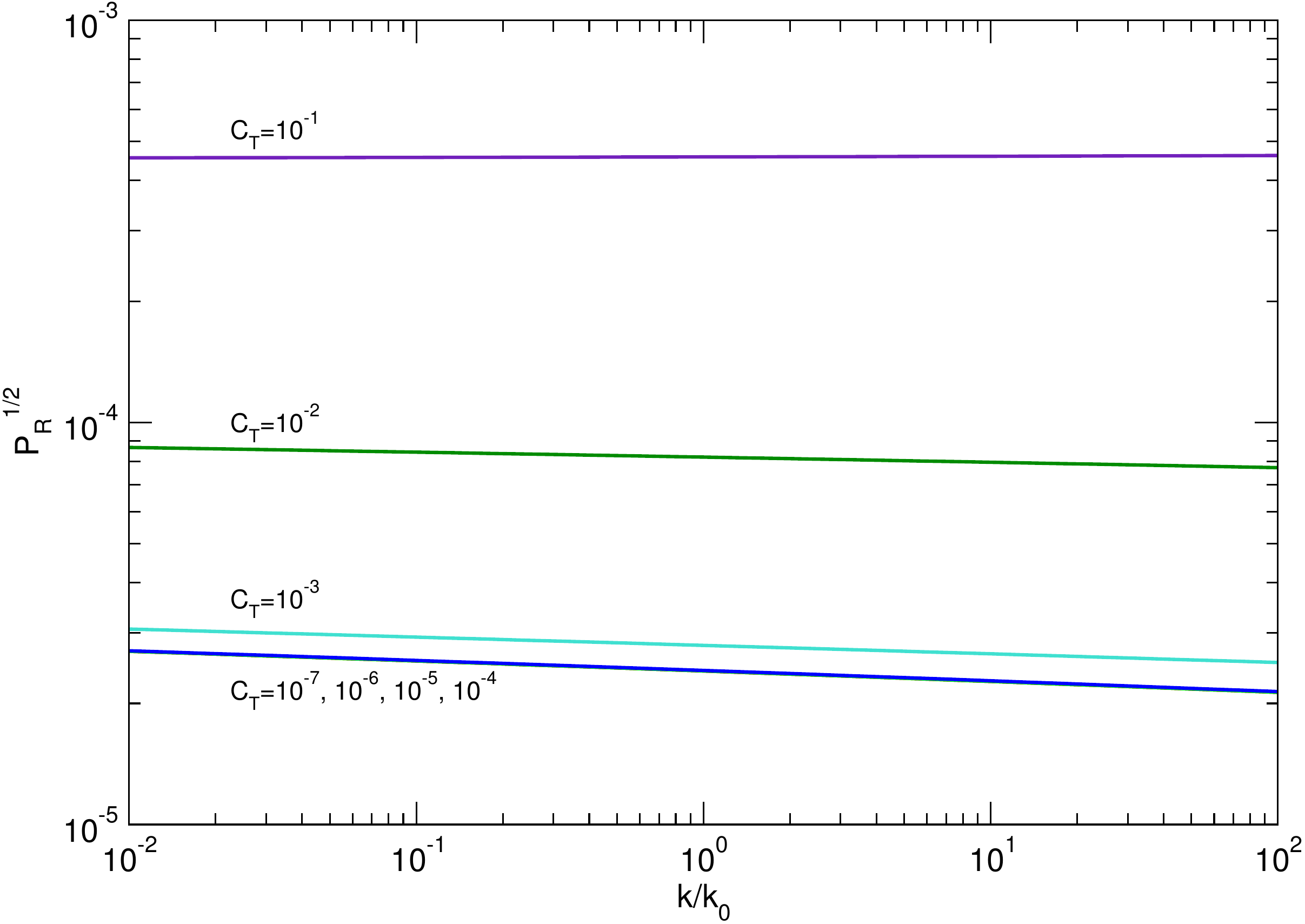}\label{f1a}}
  \hfill
  \subfloat[]{\includegraphics[width=0.50\textwidth]{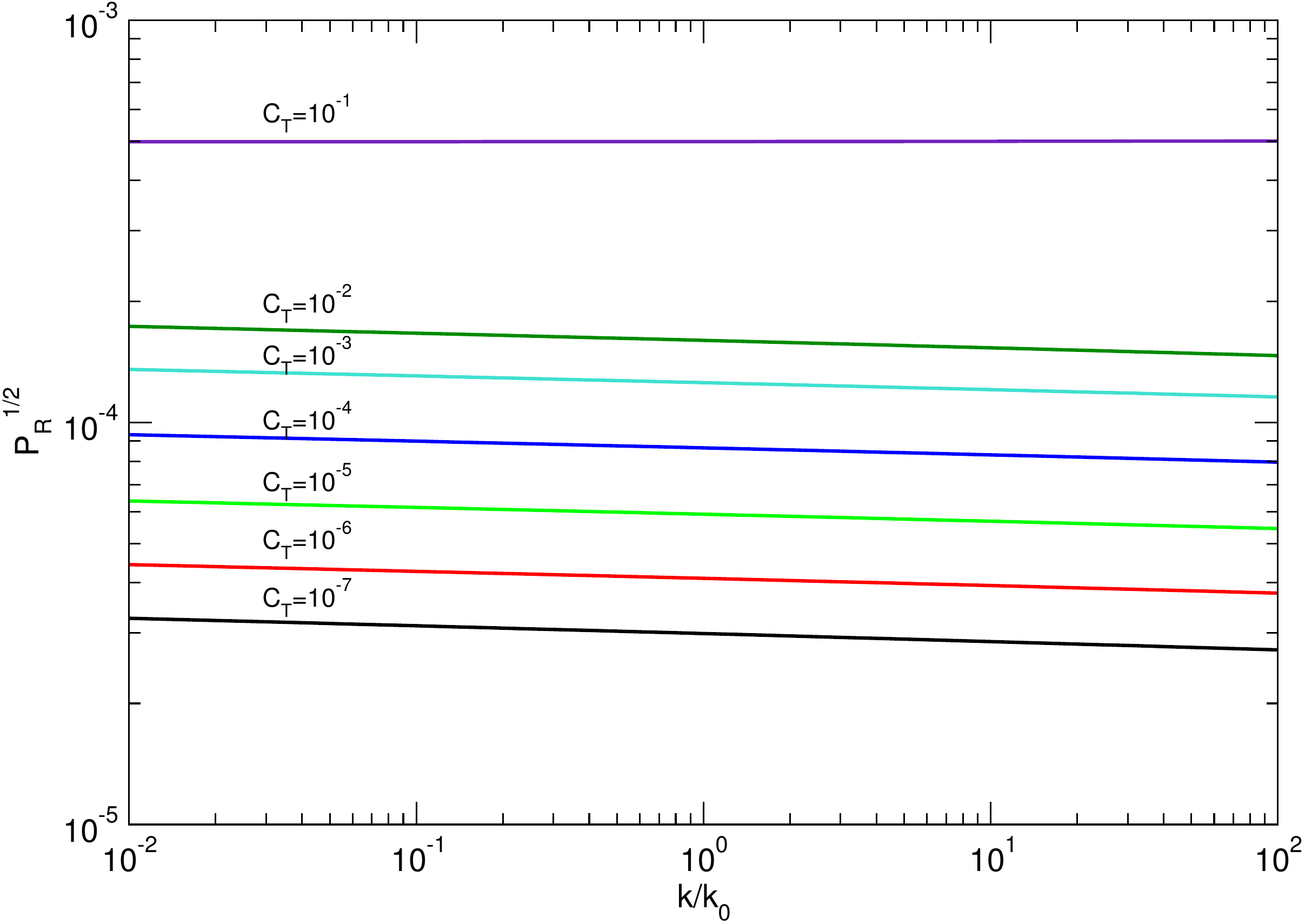}\label{f1b}}
\caption{\small{Primordial spectrum as a function of $k/k_0$, for different values of the parameter $C_T=10^{-7},\,10^{-6},\,...10^{-1}$ and for fixed $\lambda=10^{-14}$, $g_*=12.5$. Fig 1(a) is for a non-thermal inflaton, i.e, $\mathcal{N}_*=0$ and Fig 1(b) is for a thermal inflation, i.e., $\mathcal{N_*} \neq 0$.}}
\label{plot1}
\end{figure}

\begin{figure}[H]
\centering
\subfloat[]{\includegraphics[width=0.5\textwidth]{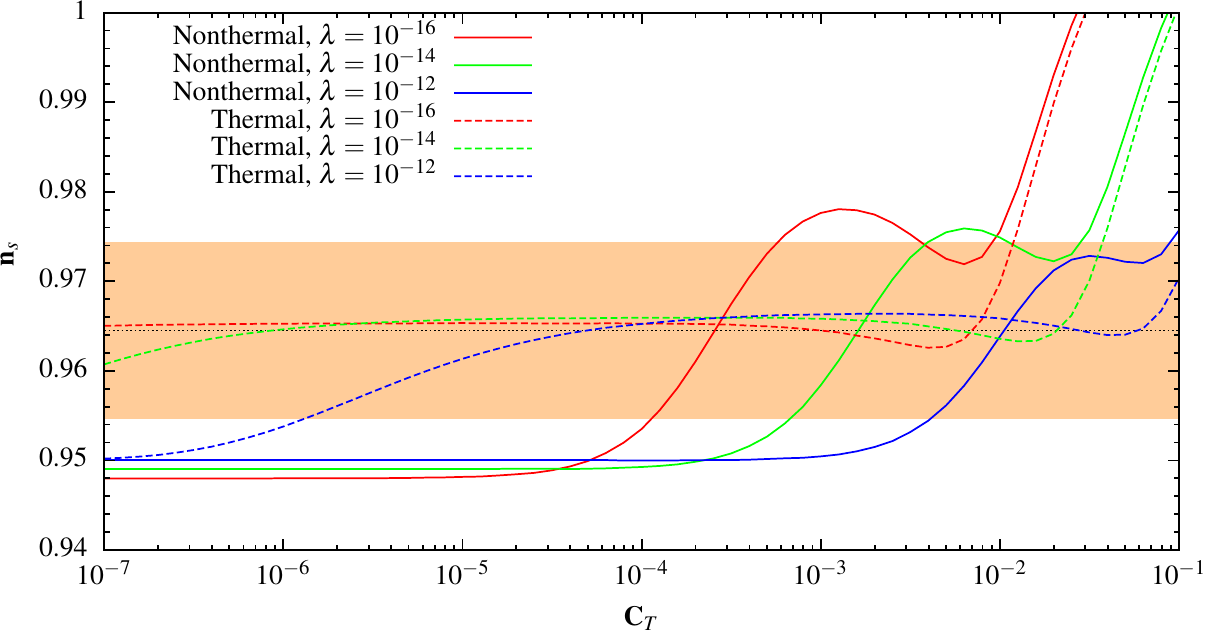}\label{f2a}}
\hfill
\subfloat[]{\includegraphics[width=0.5\textwidth]{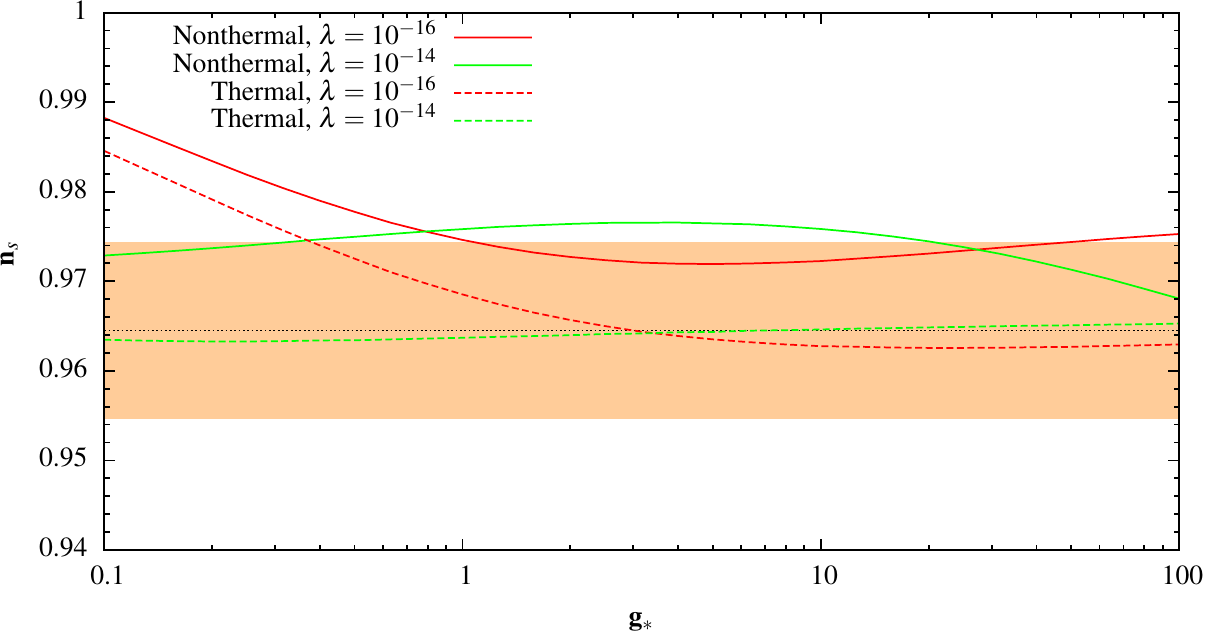}\label{f2b}}
\caption{\small{Spectral index as a function of $C_T$ with $g_*=12.5$ in Fig. (a) and as a function of $g_*$ with $C_T = 0.004$ in Fig. (b) for different values of $\lambda$ as indicated in the plot. The solid lines are for $\mathcal{N}_*=0$ and the dashed lines are for $\mathcal{N_*} \neq 0$. The horizontal black line denotes the marginalised central value for Planck TT,TE,EE+lowP data and the light brown band represents the observational 2-$\sigma$ bounds on $n_s$ from the same data combination.}}
\label{plot2}
\end{figure}

 The results for the spectrum as a function of the scale $k/k_0$ are shown in Fig. (\ref{plot1}). We have done an example for the parameter values: $\lambda=10^{-14}$, $g_*=12.5$, and different values of $C_T$ as indicated in the figure. The lowest value of $C_T$ included in the plot gives a value $Q_* \sim 10^{-7}$, whereas for $C_T \sim
10^{-1}$ , $Q_*\sim 10$. The minimum allowed value of $Q_*$ can be calculated from the condition $T_*/H_* \simeq 1$. We note that  increasing the value of $C_T$ increases the scalar amplitude, and for the non-thermal case, the amplitude reaches to an asymptotic lower value when $C_T$ becomes very small. 
To inspect the nature of the parameters better, it is judicious to compare the warm inflation power spectrum given by Eq.~\eqref{PRQ*} to the standard power law power spectrum defined as:
\be
P_{\cal R} (k) = P_{\cal R} (k_0) \left( \frac{k}{k_0} \right)^{n_s -1} \,. \label{power}
\ee
The spectral index is plotted as a function of the model parameters in Fig.~(\ref{plot2}).
The dependence is shown for three different values of $\lambda$ as indicated in the figure. In Fig.~\eqref{f2a}, the variation is over $C_T$ with $g_*=12.5$ and in Fig.~\eqref{f2b}, the variation is over $g_*$ with $C_T=0.004$. For warm inflation with $\mathcal{N}_*=0$, the $C_T$-$n_s$ plot in Fig.~\eqref{f2a} shows that for small values of $C_T \lesssim {\mathcal O}(10^{-4})$, well in the weak dissipative regime with $Q_* \ll 1$, the first term within the brackets in Eq. (\ref{PRQ*}) is negligible. Therefore, one recovers the standard expression in cold inflation where the spectrum is red-tilted and hardly depends on\footnote{The mild dependence on $\lambda$ comes from the relation between the no. of e-folds and $k$ in Eq. (\ref{Nk}).} $\lambda$. As $C_T$ ($Q_*$) increases, the dissipative contribution tends to make the spectrum less red-tilted, and for values $C_T \gtrsim 0.1$, the growing mode will render the spectrum blue-tilted. In the intermediate regime, the spectral index shows oscillatory behaviour while being roughly consistent with PLANCK $2$-$\sigma$ limits. In the case $\mathcal{N_*} \neq 0$, the spectral tilt has a little higher value than the non-thermal case for small $C_T$ due to non-zero value of $\mathcal{N_*}$ in Eq. (\ref{PRQ*}) where the contribution depends on $C_T$ as Eq. (\ref{coth}). For $C_T \gtrsim 1$, the contribution from growing mode makes the spectrum blue-tilted in a similar way as in the non-thermal case. The observational bounds on $n_s$ exclude the blue-tilt part. The Fig.~(\ref{f2b}) shows that for both non-thermal and thermal warm inflation scenarios, the variation of $n_s$ with $g_*$ is small. From this observation, we can anticipate that in the process of parameter estimation to be done in the next section, $g_*$ might not be well constrained from the limits of the spectral index.   

\begin{figure}[H]
  \centering
  \subfloat[]{\includegraphics[width=0.5\textwidth]{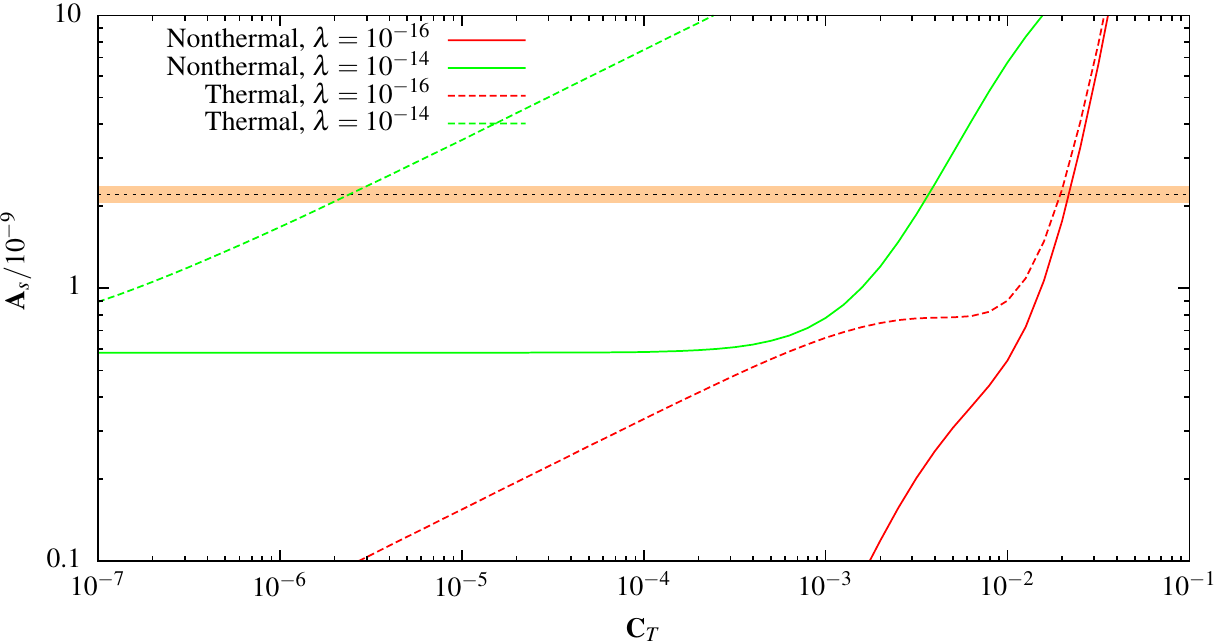}\label{f3a}}
  \hfill
  \subfloat[]{\includegraphics[width=0.5\textwidth]{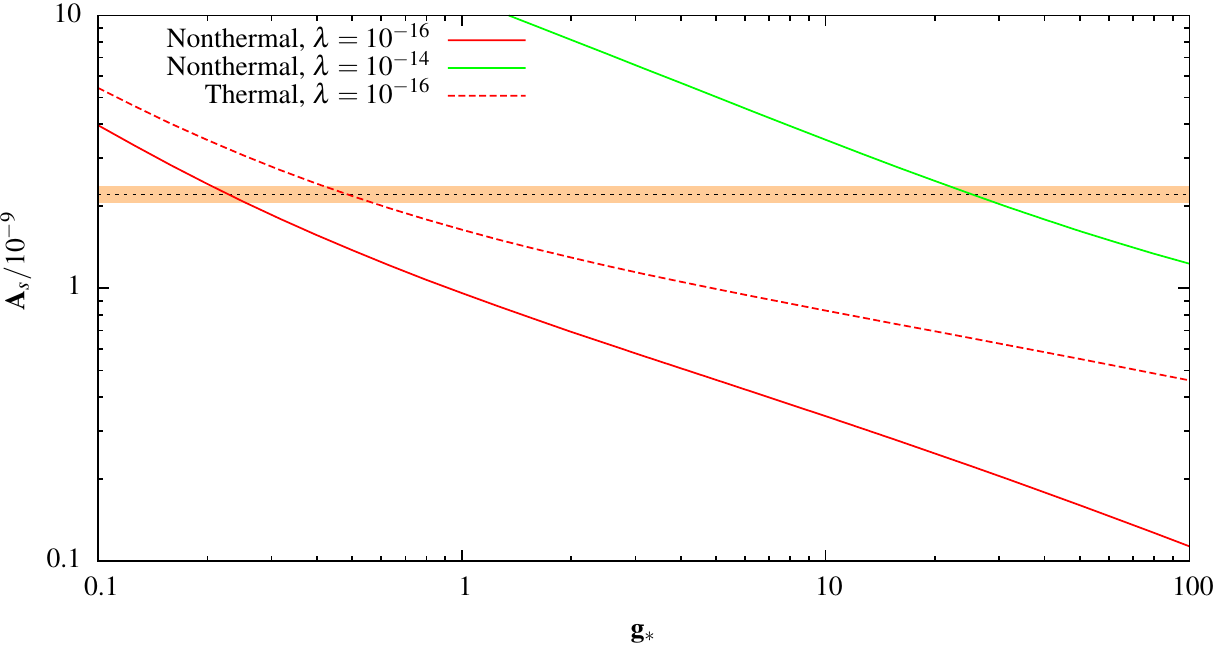}\label{f3b}}
  \caption{\small{Amplitude of spectrum $A_s$ as a function of $C_T$ with $g_*=12.5$ in Fig(a) and as a function of $g_*$ with $C_T = 0.004$ in Fig(b) for different values of $\lambda$ as indicated in the plot. The solid lines are for $\mathcal{N}_*=0$ and dashed lines are for $\mathcal{N_*} \neq 0$. The horizontal dotted black line denotes the marginalised central value for Planck TT,TE,EE+lowP data and the light brown band represents the observational 2-$\sigma$ bounds on $A_s$ from the same data combination. In Fig. (b), the thermal case with $\lambda=10^{-14}$ is not inlcuded because it gives an amplitude larger than $A_s \sim 10^{-8}$ for $g_* \lesssim 10^3$.
}}
\label{plot3}
\end{figure}

In Fig. (\ref{plot3}) we fit $P_{\cal R} (k_0) = A_s $ for the same specification of the parameters mentioned in the previous paragraph. Both Fig.~(\ref{f3a}) and Fig.~(\ref{f3b}) show that the observed range for $A_s$ allows the parameter ranges for $C_T$ and $g_*$ with tighter constraints. It is important to note that in contrast to the cold inflation scenario with a quartic potential, the amplitude of scalar perturbations in the case of warm inflation depends substantially on other parameters, namely $C_T$ and $g_*$. Analysis depicted in the Figs.~\eqref{plot2} and \eqref{plot3} helps us to choose prior ranges for the parameters to be inserted in the CosmoMC run for parameter estimations in the next section.

\begin{figure}[H]
  \centering
  \subfloat[]{\includegraphics[width=0.45\textwidth]{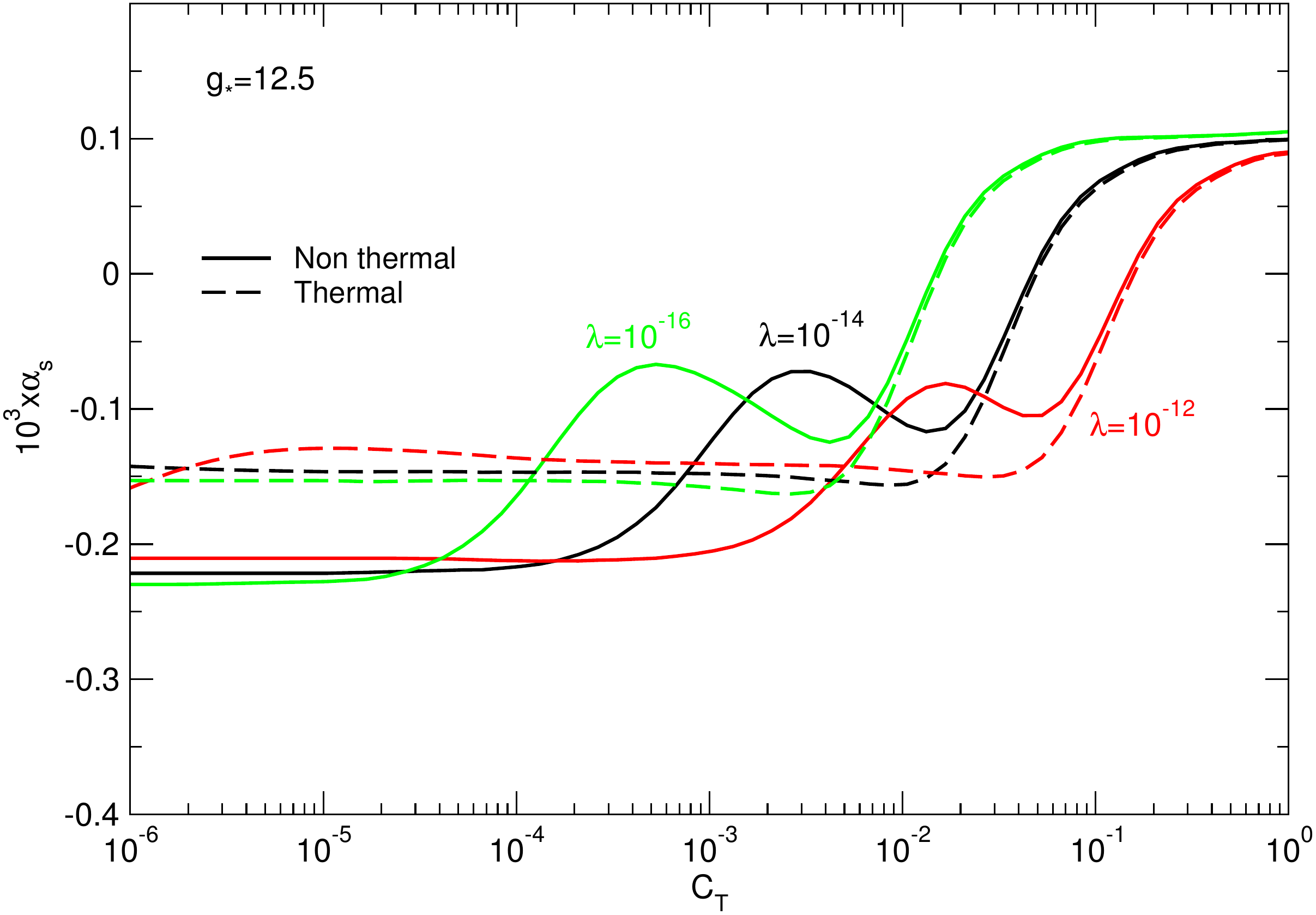}\label{f4a}}
  \hfill
  \subfloat[]{\includegraphics[width=0.45\textwidth]{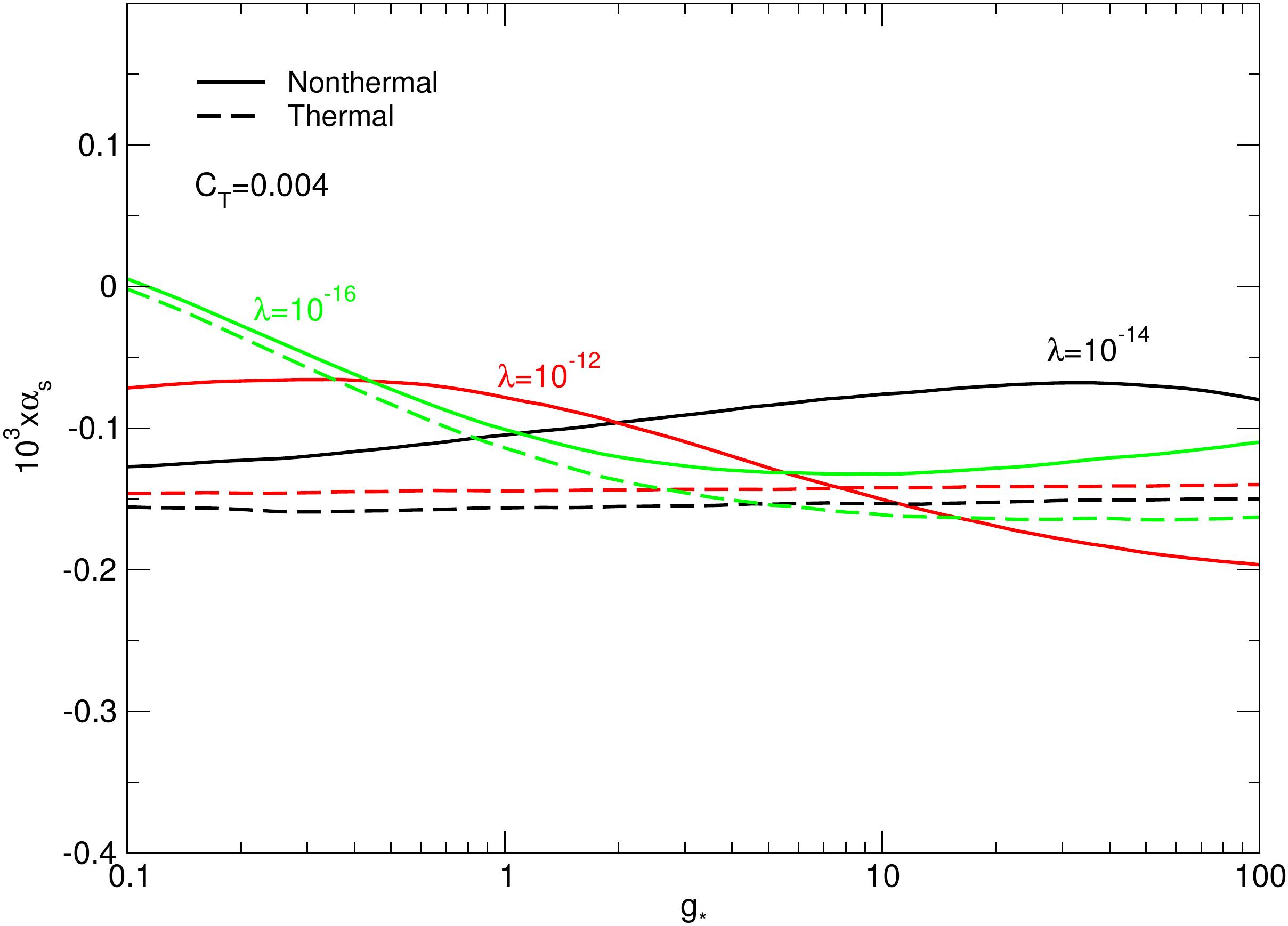}\label{f4b}}
  \caption{\small{Running of  the spectral index as a function of $C_T$ with $g_*=12.5$ in Fig. (a), and as a function of $g_*$ with $C_T=0.004$ in Fig. (b), for different values of $\lambda$ as indicated in the plot. The solid lines are for  $\mathcal{N}_*=0$ and dashed lines are for  $\mathcal{N_*} \neq 0$.}}
\label{plot4}
\end{figure}

In all the above discussions, we have neglected the running of the spectral index. Instead of a simple power law, we could include the running $(\alpha_s(k_0) = \frac{dn_s}{d(\ln k)})$ or other higher order derivatives of the spectral index in the fit as well,
\bea
P_{\cal R} (k) &=& P_{\cal R} (k_0) \left( \frac{k}{k_0} \right)^{n_s(k) -1} \,, \\
n_s(k)&=& n_s(k_0) + \frac{1}{2} \alpha_s(k_0) \ln \frac{k}{k_0} + \cdots \,. 
\eea
However, as shown in Fig.~(\ref{plot4}) this is always small with $|\alpha_s| \lesssim 10^{-4}$, as it was also found in \cite{Benetti}. And again, the change from negative to positive values of $\alpha_s$ when increasing $C_T$ is due to the growing mode. Nevertheless, given  that the estimated value of the running in this model is below the sensitivity of current and future CMB experiments \cite{Silkalphas} and assuming that higher order contributions will be lesser, we will not include the running or other higher order terms in our current analysis.  

\section{Methodology of analysis}  \label{methodology}
\paragraph*{} 

From the analysis of the previous section, we know the range for which we expect to find the best fit parameters. In our case, inflationary power spectrum, both scalar and tensor, are known in terms of three parameters: (i) $C_T$, the proportionality constant for the dissipative coefficient, (ii) $\lambda$, the quartic coupling constant for the inflaton scalar potential, and (iii) $g_*$, the total number of relativistic d.o.f in radiation bath. These three parameters can be thought equivalent to the usual parameterization by the scalar spectral amplitude $A_s$, scalar spectral index $n_s$, and tensor-to-scalar ratio $r$ representing the amplitude of tensor fluctuations. In addition to these primordial parameters, the spatially flat background cosmology is described by four other parameters, namely $\Omega_b h^2$ and $\Omega_c h^2$ ($h$ is related to the present Hubble parameter) representing baryon and dark matter densities respectively, the acoustic peak angular scale $\theta$, and the reionization optical depth $\tau$. Effectively, we have exactly the same number of parameters as like the usual $\Lambda$CDM$+r$ model. Although, our goal is to constrain the model parameters $C_T, \lambda$ and $g_*$, for convenient  comparision with the data we will quote values of $n_s$ and $r$ for the marginalised and best fit values of the parameters with the usual assumption of power spectrum given by Eq.~\eqref{power} with flat tensor spectrum. 

We analyse the warm inflation scenario using a multi-dimensional Markov Chain Monte Carlo (MCMC) simulation provided by the publicly available CosmoMC package \cite{cosmomc_main} coupled to the Planck 2015 data \cite{planck2015} and BICEP2/Keck array data \cite{keck}. This analysis uses Bayesian parameter estimation to constrain the model parameters $C_T$, $\lambda$ and $g_*$ and find respective posterior probability distributions. As outlined in the previous section, we calculate the primordial scalar and tensor spectrum for all wave vectors required by CAMB that calculate $C_l$s using the following relation \cite{christensen}:
\be
C_l = \int d(\ln k) P_{\mathcal R}(k)T^2_l(k),
\label{cltheory}
\ee
where $T_l(k)$ is the transfer function that evolves the power spectrum from the end of inflation to the last scattering surface, and it depends only on the background parameters. These $C_l$ values are fed into CosmoMC for different points in the multi-dimensional parameter space. These theoretically calculated $C_l$'s are then compared to the data using Bayesian analysis, given a prior probability distributions for the parameters that are varied. CosmoMC code analyses the parameter spaces, provides posterior probability distributions for the parameters and determines a marginalised $\chi^2$ for these distributions. We emphasise that we have modified only the inflationary sector by plugging in the power spectrum $P_{\cal R}(k/k_0)$ for the warm inflation as a function of the model parameters in stead of an usual power law expression.

\begin{figure}[H]
 \centering
 {\includegraphics[width=\textwidth , keepaspectratio ]{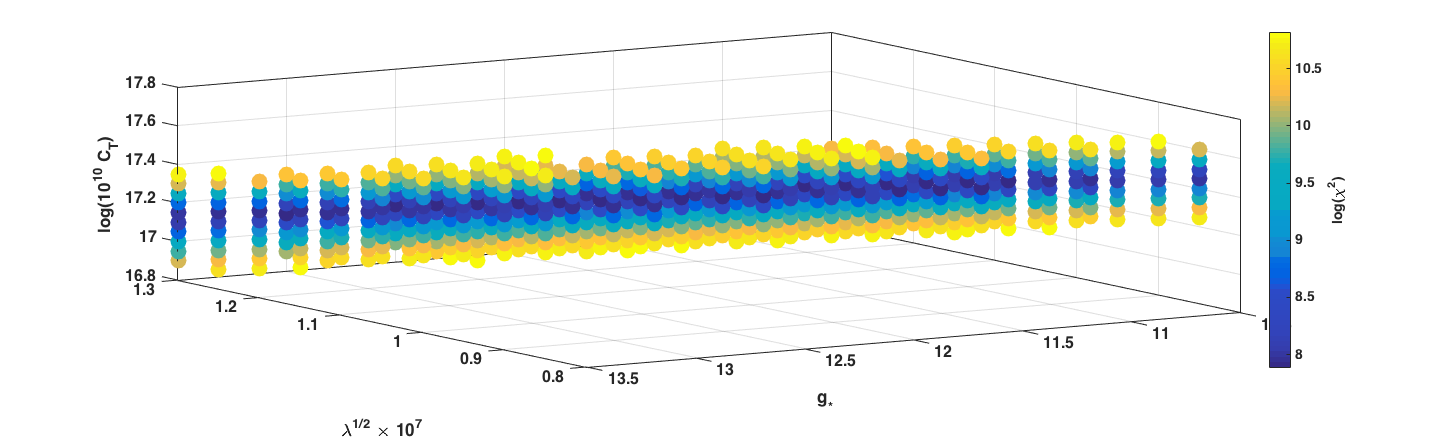}}
 \caption{\small{Scattered points in the 3-dimensional parameter space with different values of $\log(\chi^2)$ for warm inflation with non-thermal fluctuations. The points with colour in the extreme blue end of the spectrum correspond to minimum $\chi^2$. In stead of centred around a region, multiple dark blue points along a strip represent multiple modes in the probability space.}}
 \label{plot5}
\end{figure}

While doing the MCMC analysis for the warm inflation case, instead of $C_T$ we have constrained $\ln(C_T\times10^{10})$ in CosmoMC so that we can simply use the standard parameters already defined in the CAMB code. The parametrisation of $\lambda$ and $g_*$ is different for the non-thermal and thermal cases which is discussed later.
The prior ranges for the warm inflation parameters are chosen after analysing the dependence of the parameters on the pivot scalar amplitude and spectral index - See Figs. ~\eqref{plot2}, ~\eqref{plot3}. But, even then, there are multiple sets of values in the parameter space ($\lambda$, $C_T$, $g_*$) that correspond to the same value of $\chi^2$ when compared to the CMB data. We have checked it explicitly. This degeneracy is shown in Fig.~\eqref{plot5} for the case of non-thermal warm inflation ($\mathcal N_* = 0$), where the scattered points are plotted in the $3$-dimensional parameter space $\ln(C_T\times10^{10})$, $\sqrt{\lambda}\times10^7$ and $g_*$, with $\log(\chi^2)$ represented in the colour spectrum. The points with the darkest blue colour in the parameter space are the degenerate points for minimum or near-to-minimum value of $\chi^2$. The lack of clustering of these dark blue points around a single point in this plot implies that multiple degenerate points can be sampled while minimising $\chi^2$ using a typical MCMC procedure. Therefore, the posterior probability distribution of these parameters can have multiple peaks and subpeaks (multimodal systems). This was practically encountered many times while performing the MCMC analysis. Similar degeneracy can be observed in the parameter space for thermal warm inflation scenario also. For this reason, the warm inflation CosmoMC sampling faced the challenge of slow mixing and therefore slow convergence. 

For the non-thermal case ($\mathcal N_* \neq 0$), this problem was statistically dealt with the use of higher \textit{temperature} $(t)$ of the MCMC chains with the default sampling algorithm. The \textit{temperature} ($t$) defines how likely it is to sample from a low-density part of the target distribution. The advantage of low $t$ system is  more precise sampling but on the other hand it can get trapped in a local region of the phase space. Specially, in case of a theory with multiple modes, keeping low $t$ would mean definite entrapment in local modes. Though, high $t$-analysis is less precise in sampling with respect to those with low $t$, it ensures sampling of a large volume of the phase space. Thus, increasing the \textit{temperature} of the chains saved computation time without making too much compromise. The standard procedure is to set \textit{temperature} $ t=1$ in CosmoMC, and we have taken $t=2$ to serve our purpose\footnote{Note that this \textit{temperature}($t$) is merely a technical term used in the MCMC statistics and is to be distinguished from the warm inflation temperature ($T$) defined in Eq.~\eqref{deftemp}. The temperature of the chains can be changed in the common.ini file in CosmoMC. If the temperature is modified, the corresponding post-processing can be taken care of in GetDist by modifying the cool parameter.}.

For the thermal case, careful reparametrisation is needed due to the presence of non-zero ${\mathcal N}_*$ in the expression for the power spectrum given by Eq.~\eqref{PR}. Through the term ${\mathcal N}_*$, there is an overall factor of $C_T^4/C_R\lambda$. Therefore, the dependence on $g_*$ is different from the non-thermal case. For our convenience we have reparametrised $g_*$ as $19-\log(30C_T^4/\pi^2g_*\lambda)$. This reparametrisation is done following hierarchical centering \cite{gelfand} which is an algorithm to replace original parameters in a model with modified parameters that are less correlated with each other in the joint posterior distribution. The multimodality in the posterior distribution becomes more cumbersome in this case due to more mixing between the model parameters. Therefore the sampling method for the MCMC chains was also changed to Wang-Landau sampling algorithm\footnote{sampling method=6 in the CosmoMC package} \cite{wang1,wang2} which is better sampling to tackle unknown target distributions. In addition, the \textit{temperature} of the chains is also increased to 2. All these statistical tweaks help to deal with the secondary peaks and long tails in the posterior distributions and lead to faster and better convergence as well.
  


\section{Results and Discussions}
\paragraph*{}
\label{results}

In this section we present our results both for the thermal and non-thermal case. The CosmoMC code constrains the model parameters as well as the late time cosmological parameters for warm inflation and estimates the posterior probability distribution with marginalised central values and standard deviations. 

The following Figs.~\ref{plot6} and \ref{plot7} show the posterior distributions for the model parameters. The parametrisation is done as ($\ln(C_T\times10^{10})$, $\sqrt{\lambda}\times10^7$, $g_*$) for the non-thermal case (Fig.~\ref{plot6}) and as ($\ln(C_T\times10^{10})$, $\sqrt{\lambda}\times10^7$, $19-\log(30C_T^4/\pi^2g_*\lambda)$) for the thermal case (Fig.~\ref{plot7}) respectively as mentioned in the earlier section. The likelihoods used here are \textit{Planck TT+TE+EE}, \textit{Planck lowP}, estimated using commander, Planck lensing and \textit{BICEP2$/$Keck array and Planck} joint analysis likelihood\cite{keck,commander,bkplanck}. These plots show both one-dimensional and two-dimensional marginalised posterior distributions for these parameters. The marginalised central values are determined by post-processing using Getdist package included in CosmoMC. \\
\begin{figure}[H]
\centering
{\includegraphics[width=9cm, height = 9cm ]{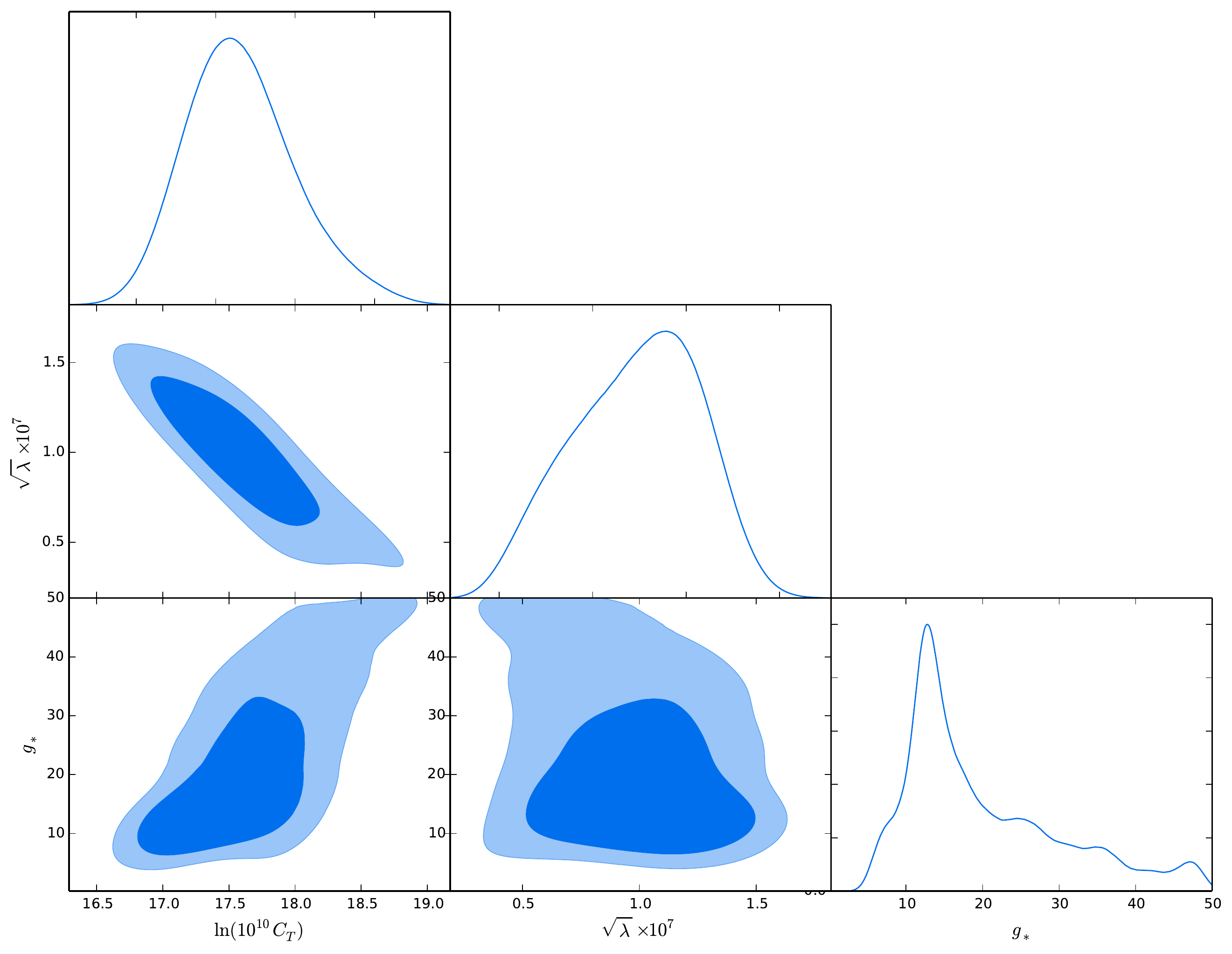}}
\caption{\small{Triangle plot for the model parameters $C_T$, $\lambda$ and $g_*$ when $\mathcal{N}_*=0$. Diagonal plots are the marginalised probability densities for these parameters and off diagonal plots represent $68\%$ and $95\%$ confidence limits for the variation of two sets of model parameters.}}
\label{plot6}
\end{figure}
\begin{table}[h!]
\centering
\caption{Constraints on cosmological parameters for non-thermal and thermal case compared with $\Lambda CDM +r$ using Planck 2015+BICEP2/Keck Array\cite{keck,commander,bkplanck} observations .}
\resizebox{\textwidth}{!}{
\begin{tabular}{|c|c|c|c|c|c|c|c|}
\hline
\multicolumn{5}{|c|}{Warm Inflation} & \multicolumn{3}{|c|}{Cold Inflation}\\
\hline 
 & \multicolumn{2}{|c|}{$\mathcal{N}_*=0$} & \multicolumn{2}{|c|}{$\mathcal{N}_* \neq 0$}& & \multicolumn{2}{|c|}{$\Lambda CDM+r$} \\ 
\hline
parameters & mean value & $1\sigma$ & mean value & $1\sigma$ & parameters & mean value & $1\sigma$\\
\hline 
$\Omega_bh^2$ & 0.02233 & 0.00022 & 0.02224 & 0.00019 & $\Omega_bh^2$ & 0.02224 & 0.00017\\ 
\hline 
$\Omega_ch^2$ & 0.1178 & 0.0015 & 0.1194 & 0.0013 & $\Omega_ch^2$ & 0.1192 & 0.0016 \\
\hline 
$100\theta_{MC}$ & 1.04097 & 0.00046 & 1.04088 & 0.00038 & $100\theta_{MC}$ & 1.04085 & 0.00034\\
\hline 
$\tau$ &  0.077 & 0.019 & 0.068 & 0.021 & $\tau$ & 0.064 & 0.018\\
\hline 
$C_T$ & 0.0043 & 0.0018 & 0.0104 & 0.0077 & $\ln(A_s\times10^{10})$ & 3.06 & 0.031\\
\hline 
$\lambda$ & 9.77$\times 10^{-15}$ & 5.41$\times 10^{-15}$ & 9.74$\times 10^{-16}$& 6.78$\times 10^{-16}$ & $n_s$ & 0.966 & 0.0052\\
\hline 
$g_*$ & 20.03 & 10.39 & 139.91 & 487.98 & $r$ & \multicolumn{2}{c|}{$<$ 0.07}\\
\hline
\end{tabular}}
\label{marginalised1}
\end{table}

In Table (\ref{marginalised1}), the marginalised values for the model parameters along with the late time cosmological parameters in $\Lambda$CDM model are quoted with their respective 1-$\sigma$ errors. In the case of the non-thermal warm inflation, it can be seen from Fig.~\ref{plot6} that the posterior probability for $g_*$ has a long tail and is far from a Gaussian distribution. This can be interpreted as an effect of the degeneracy in the parameter space as mentioned earlier in Sec. \ref{methodology}. Therefore, the marginalised mean value and standard deviation for $g_*$ in Table (\ref{marginalised1}) is not completely conclusive as marginalisation is done by fitting the posterior distribution as a Gaussian. The marginalised mean value for the inflationary parameters are as follows: $C_T = 0.0043$, $\lambda = 9.77\times 10^{-15}$, and $g_* = 20.03$.  Looking at the $1$-$\sigma$ values, the current set of data along with the default algorithm \cite{metropolis} that we use for the MCMC analysis cannot constrain the $g_*$ parameter stringently. On the other hand $C_T$ and $\lambda$ are well constrained. To have a better understanding of this, we also quote the best-fit values of the warm inflation parameters for the non-thermal case: $\lambda \sim 1.38\times 10^{-14}$, $C_T \sim 0.0030$, $g_* \sim 12.32$. It is intersting to note that the most likely value of $g_*$ is close to the particle content proposed in the model in Ref.~\cite{BBRR16}. These values denote the positions of the maximum posterior probability in the triangle plot of Fig.~\ref{plot6}. We note that for the case of $g_*$, the position of the maximum posterior probability and the mean value from the marginalised one-dimensional plot differs by $1$-$\sigma$. The marginalised mean values for other background cosmological parameters are consistent upto $1$-$\sigma$ confidence level for $\Lambda$CDM$+r$ model for the same data combinations - See Table (\ref{marginalised1}). 

\begin{figure}[H]
  \centering
  \subfloat[]{\includegraphics[width=0.45\textwidth]{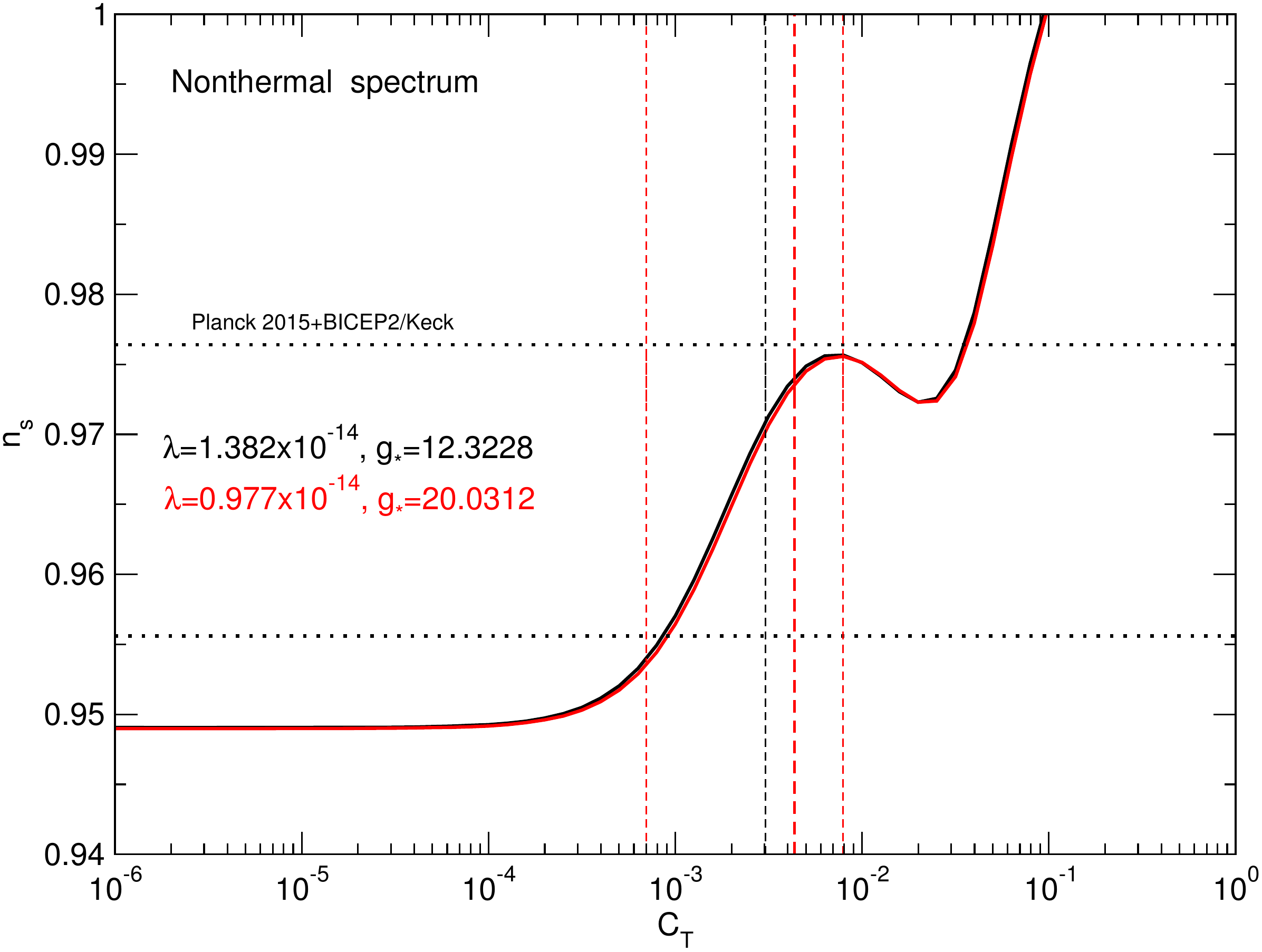}\label{f7a}}
  \hfill
  \subfloat[]{\includegraphics[width=0.45\textwidth]{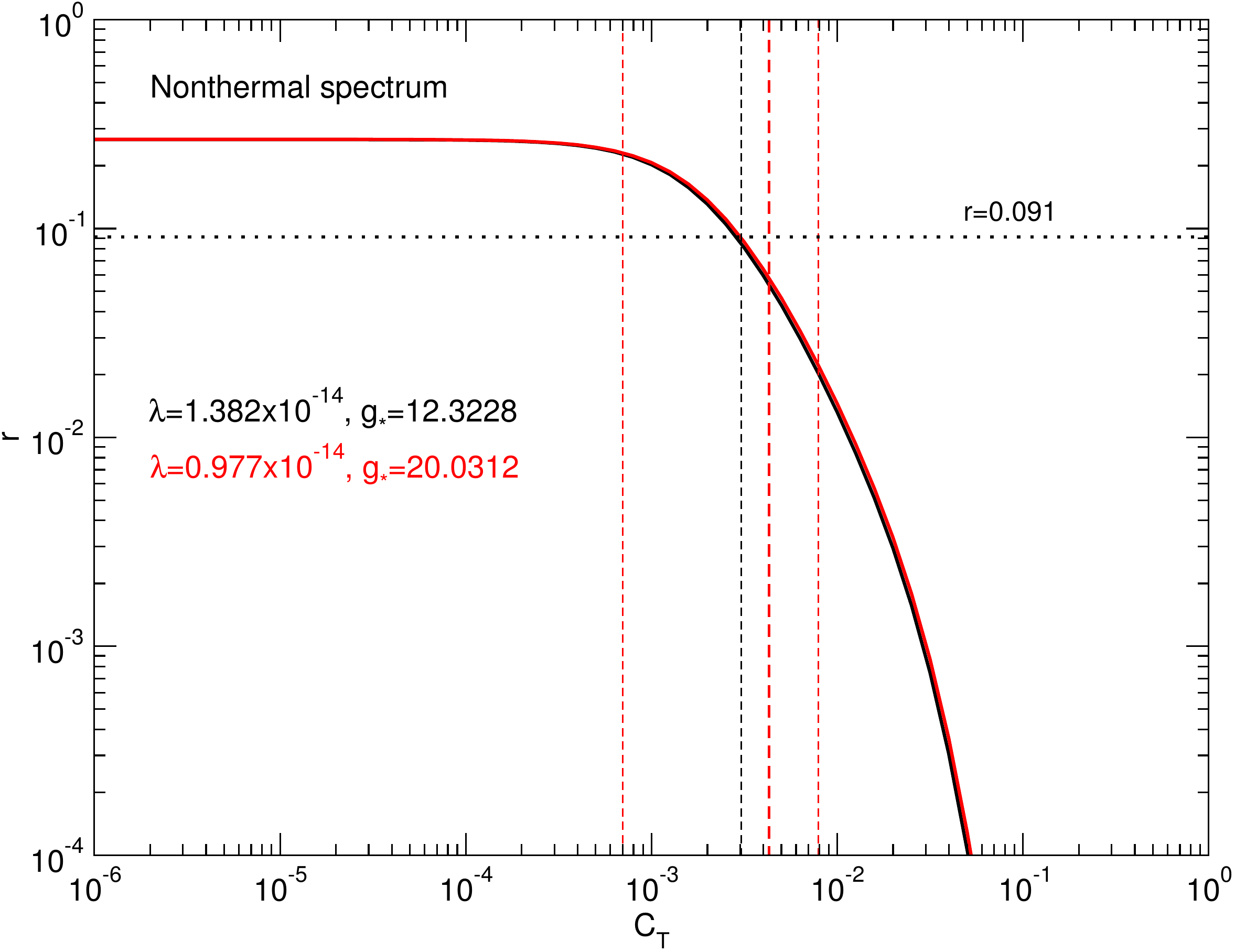}\label{f7b}}
  \caption{\small{The predictions for the spectral index and tensor-to-scalar ratio for the best-fit (black) and mean value (red) of parameters for non-thermal case. The vertical black dotted line corresponds to the best-fit value of $C_T$, whereas dotted red lines corresponds to the mean value (central), and its $2$-$\sigma$ limit as given in Table. \ref{marginalised1}. In Fig. (a), the horizontal lines correspond to the $2$-$\sigma$ constraints for different data combinations, whereas horizontal line in Fig. (b) corresponds to the current upper limit on $r$. }}
\label{plot10}
\end{figure}

To understand the consistency of the model, it is instructive to find the inflationary observables for the best-fit parameters and compare those with bounds from the recent data. We show this in Fig.~\ref{plot10} where we plot the scalar spectral index $n_s$ and the tensor-to-scalar ratio $r$ as a function of $C_T$ for both the best-fit parameter and marginalised mean values for $\lambda$ and $g_*$, and those are not much different from each other. The best-fit parameters are $n_s = 0.9709$, $r = 0.09$ with running $\alpha_s = -6.7\times 10^{-5}$, whereas parameters for the marginalised mean values are $n_s = 0.9736$, $r = 0.06$ with $\alpha_s = -7.2\times 10^{-5}$. For the marginalised mean value of the parameters, we find $Q_* = 0.031$ with $T/H_* = 21.3$, and the pivot scale exits the horizon $N_* = 58$ e-folds before the end of inflation, whereas for the case of best-fit values, we find $Q_* = 0.019$ with $T/H_* = 19.3$, and the horizon exit scale happened $N_* = 58$. As we have argued earlier, the running is always negligible. The vertical lines in Fig.~\ref{plot10} corresponds to the mean value for $C_T$ and their $2$-$\sigma$ error bars, and the best fit value. The horizontal lines corresponds to the observational constraints. We see that smaller values of $C_T \lesssim 10^{-3}$ are excluded as it predicts larger tensor amplitude and too small scalar tilt. $C_T$ in the range of $10^{-3}$ and $10^{-2}$ could have been consistent with both the constraints from $n_s$ and $r$, but in that range, it predicts too large scalar amplitude - see Fig.~\ref{f3a}. Therefore, we see that for the non-thermal case, the preferred values of the parameters predict $r$ that is close to the current upper limit, and further constraint on $r$ would either validate or exclude the set-up. In particular, non-observation of $r \sim 0.01$ would strongly constrain the scenario.

\begin{figure}[H]
\centering
{\includegraphics[width=9cm, height = 9cm ]{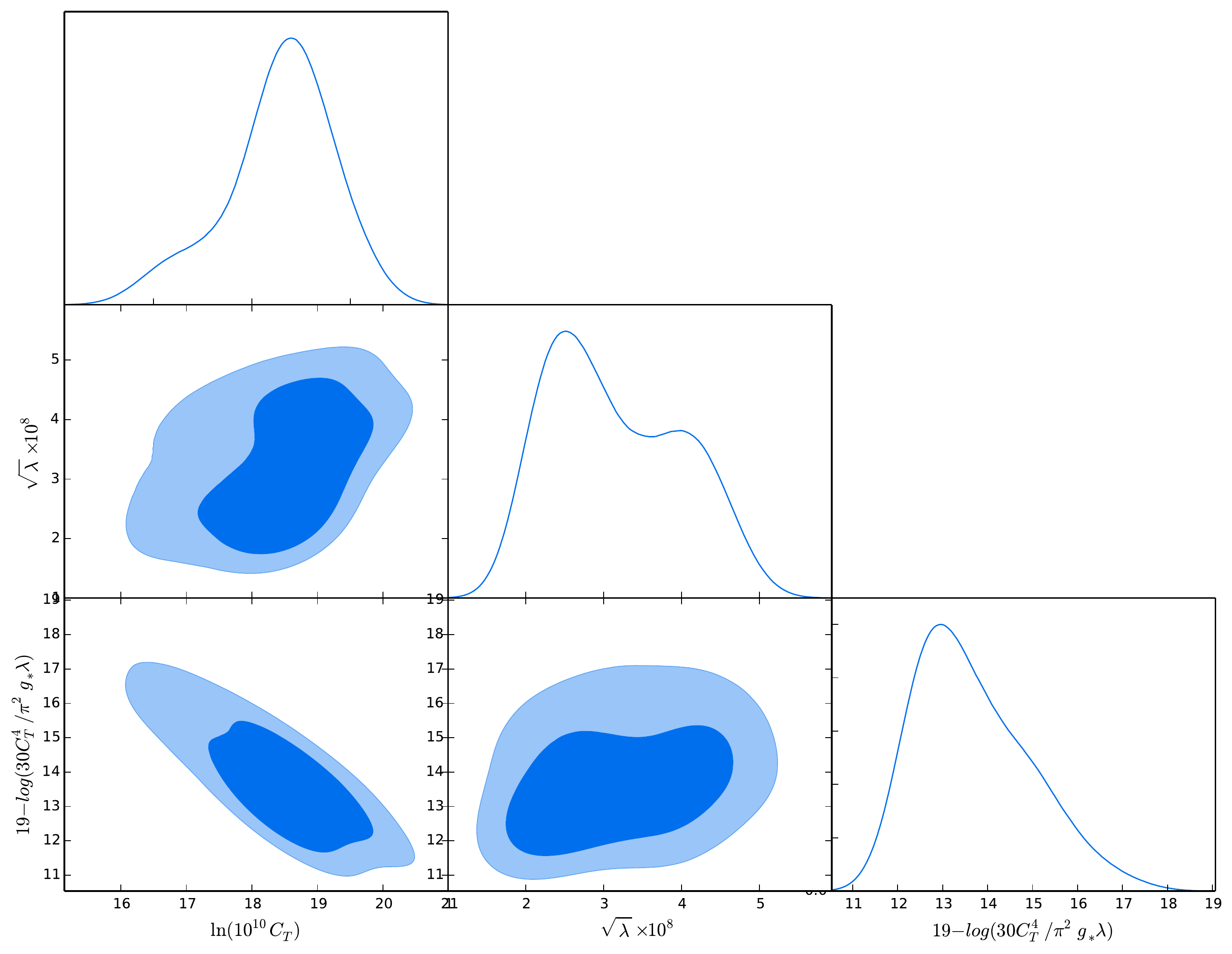}}
\caption{\small{Triangle plot for the model parameters $C_T$, $\lambda$ and $g_*$ when $\mathcal{N}_*\neq 0$. Diagonal plots are the probability density for these parameters and off diagonal plots represent $68\%$ and $95\%$ confidence limits for the variation of two sets of model parameters}}
\label{plot7}
\end{figure}

Now, we turn to the discussion of thermal case with ${\mathcal N}_* \neq 0$. In thermal warm inflation scenario, the marginalised mean values of the parameters are (see Table \ref{marginalised1}): $C_T = 0.0104$, $\lambda = 9.74\times 10^{-16}$, and $g_* = 139.91$. Here, we note that the observations are unable to tightly constrain the number of thermal degrees of freedom $g_*$ as compared to the non-thermal case, and this was anticipated from the analyses in Sec. \ref{back_ana}. Both $C_T$ and $\lambda$ values are well constrained in this case, and $C_T$ is larger by one order of magnitude compared to the non-thermal case, whereas $\lambda$ is smaller by a similar amount. We also quote the best-fit values of the parameters here: $C_T = 0.0032$, $\lambda = 9.6145\times 10^{-16}$, and $g_* = 126.7637$. The marginalised mean values of the background cosmological parameters are consistent with those from the $\Lambda$CDM+r run upto $1-\sigma$ confidence level for the same set of data combinations.

\begin{figure}[H]
  \centering
  \subfloat[]{\includegraphics[width=0.45\textwidth]{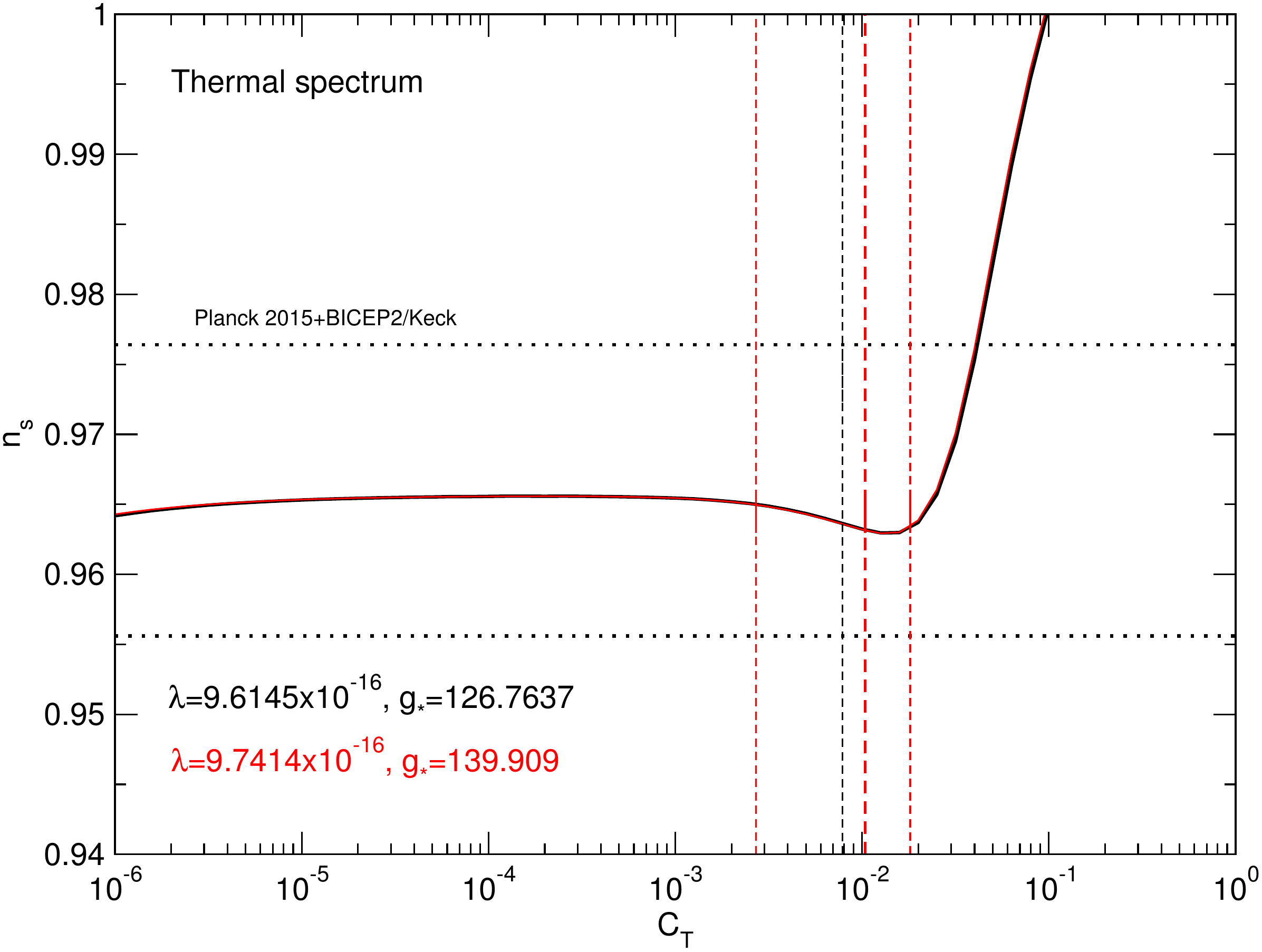}\label{f7a}}
  \hfill
  \subfloat[]{\includegraphics[width=0.45\textwidth]{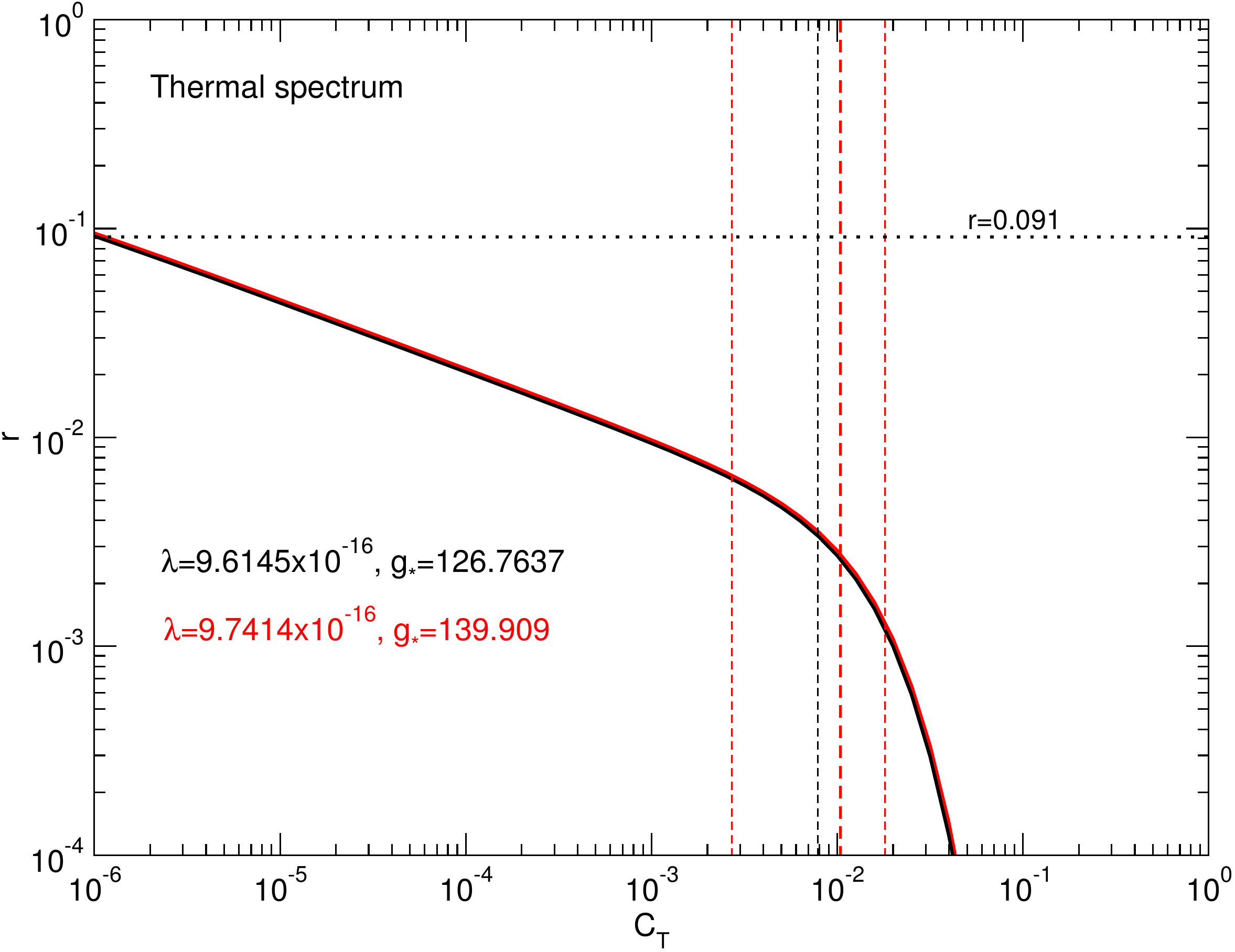}\label{f7b}}
  \caption{\small{The predictions for the spectral index and tensor-to-scalar ratio for the best-fit (black) and mean value (red) of parameters for thermal case. The vertical black dotted line corresponds to the best-fit value of $C_T$, whereas dotted red lines corresponds to the mean value (central), and its $1$-$\sigma$ limit as given in Table. \ref{marginalised1}. In Fig. (a), the horizontal lines correspond to the $2$-$\sigma$ constraints for different data combinations, whereas horizontal line in Fig. (b) corresponds to the current upper limit on $r$. }}
\label{plot11}
\end{figure}

We also find the inflationary observables for the mean and best-fit values of the model parameters for comparison with recent observations. Fig.~\ref{plot11} shows $n_s$ and $r$ as a function of $C_T$ for mean and best-fit values of $\lambda$ and $g_*$, and these two curves are almost overlaps. The vertical dashed red and black lines correspond to the mean and best-fit values of $C_T$ whereas the thin red dotted lines correspond to the $2$-$\sigma$ error in $C_T$. Horizontal lines are the bounds from recent Planck observations. The marginalised mean values of the parameters predict $n_s = 0.9631$, $r = 0.03$ with running $\alpha_s = -1.6\times 10^{-4}$ whereas, for the best-fit values of the parameters, the observables are $n_s = 0.9648$, $r = 0.06$ with running $\alpha_s = -1.6\times 10^{-4}$. The running is very small as discussed earlier. The marginalised mean values of the parameters predict $Q_* = 0.14$ with $T/H_* = 40.70$ and $N_*=58.5$ for the horizon exit of the pivot scale. The best-fit values of the parameters give $Q_* = 0.24$ with $T/H_* = 22.5$ and $N_*=58.07$. We note that the thermal scenario predicts lower values of the tensor-to-scalar ratio $r$ than that predicted in non-thermal warm inflation case and $r$ for thermal warm inflation is well within the bounds of the present observations.

\begin{figure}[H]
\centering
{\includegraphics[width=11cm, height=6cm ]{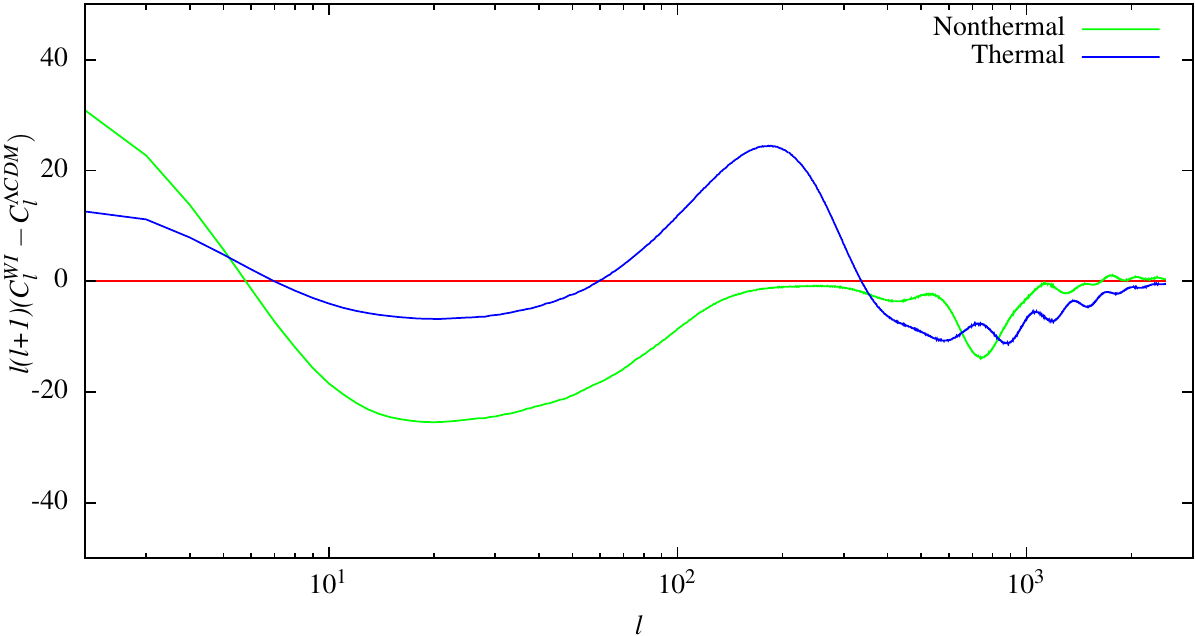}}
\caption{\small{Temperature power spectrum residual plots for the best-fit values of the model parameters for both non-thermal (green) and thermal (blue) cases with respect to the $\Lambda$CDM+r model for the data combination same as Table~\ref{marginalised1}}}
\label{plot12}
\end{figure}

In Fig.~\ref{plot12}, the difference in the temperature power spectrum for non-thermal and thermal warm inflation cases with the $\Lambda$CDM+r model is plotted for the best-fit values of the model parameters quoted above in the corresponding cases for the data combination Planck 2015+BICEP2/Keck Array. 

Although Ref.~\cite{Benetti} did a similar MCMC analysis, there are considerable diferences between our methodology of analysis with respect to theirs. Here, the complete power spectra $P_{\mathcal{R}}$ and $P_T$ are calculated numerically using Eq.~\eqref{PRQ*} and Eq.~\eqref{PTparam} while feeding them inside CAMB rather than a power-law fitting approximation in~\cite{Benetti}. Moreover,~\cite{Benetti} used Bound Optimization BY Quadratic Approximation (BOBYQA) algorithm which is not the case in our MCMC methodology. Finally, $n_s$ is calculated as a function of the marginalised (and best-fit) values of the model parameters and corresponding $r$ values are also quoted in this section which is different from the approach in~\cite{Benetti} where all the observables are calculated for $n_s=0.9655$, the mean value from \textit{Planck TT+low P}~\cite{planck2015cosmo}. Therefore, the values of $Q_*$ at the pivot scale are different in our case than that mentioned in~\cite{Benetti}. This explains the difference in observables such as $r$ and $\alpha_s$ in our thermal case from that of quartic potential with linear dissipation in~\cite{Benetti}.

The difference in the mean (and best-fit) values of $g_*$ for non-thermal and thermal cases is due to difference in effective thermalisation in these two cases. The minimum number of d.o.f. required to get this kind of linear dissipation is $g_*=11.5$ for the non-thermal case. For the thermal case, the inflaton, having a thermal (BE) distribution, contributes to $g_*$ and increases the minimum required $g_*$ to be 12.5 ~\cite{BBRR16}. Whether or not SM or BSM fields are present in the thermal bath depend on how the inflaton+dissipation sector couples to those fields. Incomplete thermalization of the (B)SM fields due to weak coupling can result in a suppressed value of effectively thermalized d.o.f. $g_*$ ($<O(100)$). The marginalised values of $g_*$ from Table~\ref{marginalised1} imply that for a warm non-thermal inflation, the preference is for a thermal bath made of the dissipative sector, but not yet thermalised (B)SM sector; whereas, for warm thermal inflation, more d.o.f. are included in the thermal bath than the minimal sector. Whether those are SM or BSM fields is a question of model building.

It is worth mentioning here, there are some of the other features in the observables that can help to distinguish between WI from CI. One feature as discussed in~\cite{Benetti} is the sign of the running of the running of the spectral index ($\beta$). The recent observations by Planck~\cite{planck2015} hint that $\beta$ ($\beta = 0.025\pm0.013$) could be positive, which contradicts the expectation from the standard CI models, whereas a quartic potential in WI scenario can predict a positive $\beta$. Another way to distinguish is by studying the non-Gaussianities as WI have some distinct features in the shapes of the bispectrum when compared to the CI picture \cite{nong-mar}.

\section{Conclusions}
\label{conclusions}

In this work we have explored the possibility of constraining parameters of the warm inflationary scenario when comparing its predictions directly with the latest CMB data. Warm inflation just takes into account possible dissipative effects induced by the interactions of the inflaton field with other species; interactions needed anyway in order to be able to reheat the universe after inflation ends. But given the variety of possibilities when combining inflationary models with patterns of dissipation, we have chosen to work (a) with a simple model of inflation, a chaotic model with a quartic coupling $\lambda$; (b) a linear $T$ dissipative coefficient, given by the interactions of a few fermions and scalars with the inflaton. Consistency of the model with observations has been already established studying its background dynamics and the primordial spectrum \cite{BBRR16}, by varying the parameters of the model. Indeed, for chaotic models dissipation helps in sustaining inflation for longer, lowering  the value of potential at horizon crossing, and therefore the tensor-to-scalar ratio. 

We have used CosmoMC to get the parameter estimation. As parameters of the model we have: the combination of coupling constants giving rise to dissipation, $C_T$ in Eq. (\ref{CT}), the effective number of relativistic degrees of freedom contributing to the thermal bath $g_*$, and the quartic coupling in the inflaton potential $\lambda$. We work directly with the scale dependent primordial spectrum, $P_{\cal R}(k)$. In principle, the calculation of the primordial spectrum is done as a function of the no. of e-folds, Eq. (\ref{PRQ*}), and to get the relation with the scale $k$ one needs to assume something about reheating: at least an effective equation of state during reheating $\tilde w$, and how long it will take for the universe to become radiation dominated.
However, these extra assumptions are avoided in our case given that (a) we already have radiation produced during inflation, (b) the quartic chaotic model behaves as radiation once the field start oscillating after inflation so that $\tilde w=1/3$. Therefore we have used Eq. (\ref{Nk}) to convert the $N$ dependence into $k$-dependence. 

In warm inflation, the presence of the thermal bath and its fluctuations can affect also the statistical state of the inflaton fluctuations, and this is taken into account with the term ${\cal N}_*$ in Eq. (\ref{PRQ*}). We have considered in our analyses that either the inflaton remains in its standard Bunch-Davies vacuum, with ${\cal N}_*=0$ (nonthermal case), or that it is in a thermal excited state with  ${\cal N}_*=n_{BE}$ (thermal case). The main results of our analyses are given in Fig. (\ref{plot6}) for the nonthermal case and Fig. (\ref{plot7}) for the thermal case. The constraints on the model parameters are given in Table \ref{marginalised1}. Notice that we have the same no. of parameters in our analyses than in the standard cold inflation one, but we have traded the three parameters related to primordial spectrum with out model parametrisation: $C_T$, $\lambda$ and $g_*$. When studying the parameter dependence of the observables in Sect. 4 we checked that in this model the running of the spectral index is always small, $|\alpha_s| \lesssim O(10^{-4})$, so the power spectrum can be well fitted by a simple power law.

In both cases, thermal and non-thermal, the less constrained parameter is the effective no. of relativistic degrees of freedom $g_*$; mainly because this parameter always appears in the combination $C_T^4/(\lambda C_R)$, with $C_R=\pi^2g_*/30$. Still, in the non-thermal case the behaviour of the spectral index with $C_T$ (see Fig. (\ref{plot2}.a)) selects a preferred range for this value, the amplitude of the spectrum do the same for $\lambda$, and therefore $g_* \simeq O(20)$. This is of the same order as the minimum no. of degrees of freedom needed to get this kind of linear dissipation, $g_*=12.5$ \cite{BBRR16}. The Monte Carlo analyses indeed returns values for the parameters such that the spectral index is as close as possible to the $\Lambda$CDM + Cold Inflation analyses mean value, i.e., $n_s\simeq 0.966$, which in this case implies a slightly larger value for the tensor-to-scalar ratio than in Cold Inflation, close to the upper limit. Future data providing a more restrictive upper limit on $r$ may then disfavor this scenario.

In the thermal case the problem of the degeneracy among the parameters is stronger. In addition, the behavior of the spectral index with $C_T$ is rather flat (until it does increase owing to the growing mode), as can be seen in Fig. (\ref{plot2}.a), which does not help with the parameter estimation. Because of that we have explored different parametrisations and method in order to get the best possible estimation. Still, we hardly get any constraint on $g_*$. Nevertheless, the typical value for the spectral index is again very close to the $\Lambda$CDM + CI, whereas in this scenario the tensor-to-scalar ratio is further suppressed with $r \simeq 0.006$. Still, it may be within the range of next generation CMB experiments.

\subsubsection*{Acknowledgements}
The authors would like to thank the organisers of {\it III Saha Theory Workshops: Aspects of Early Universe Cosmology} where the work was initiated. MBG would like to thank Theory Division of Saha Institute of Nuclear Physics for hospitality during the early stages of the project. SB and KD would like to thank University of Granada for hospitality during the final stages of the project. We would like to sincerely thank Dhiraj Kumar Hazra, Kiyotomo Ichiki, Anthony Lewis and Jayanti Prasad for valuable suggestions related to CosmoMC. MRG and SB would like to thank the Cosmocoffe discussion forum for many valuable inputs. SB is supported by a fellowship from CSIR, Govt of India.
KD is partially supported by a Ramanujan Fellowship funded by SERB, DST, Govt. of India. \par
\textbf{Note:} Right after this manuscript appeared in arXiv, Ref.~\cite{raghuwarm} was submitted in arXiv. They too carried out a CosmoMC analysis of warm inflation. There, they considered $\Upsilon \sim T^3$, whereas we concentrated on the linear dissipative regime.




\end{document}